\documentclass[twocolumn,tighten,usenames,dvipsnames]{aastex62}

\usepackage[normalem]{ulem}
\usepackage{color}
\definecolor{epcol}{rgb}{0.398, 0.0, 0.797}
\def\del#1{{}}

\usepackage{graphicx}
\usepackage{amsmath}
\usepackage{amssymb}
\usepackage{xspace}
\usepackage[normalem]{ulem}
\usepackage{mathrsfs}
\usepackage{longtable}

\usepackage{natbib}

\def\eV{{\rm eV}} 
\def\GeV{{\rm G}\eV} 
\def\TeV{{\rm T}\eV} 

\def\G{{\rm G}} 

\def\Fermi{{\em Fermi}\xspace}
\def\P{\mathcal{P}}

\def\apjs{Astrophys. J. Suppl. Ser.}
\def\apjl{Astrophys. J. Lett.}

\def\aap{Astron. Astrophys.}

\begin{document}

\title{Constraints on the Intergalactic Magnetic Field from Bow Ties in the Gamma-ray Sky}

\author[0000-0003-3826-5648]{Paul Tiede}
\affiliation{Department of Physics and Astronomy, University of Waterloo, 200 University Avenue West, Waterloo, ON, N2L 3G1, Canada}
\affiliation{ Waterloo Centre for Astrophysics, University of Waterloo, Waterloo, ON N2L 3G1 Canada}
\affiliation{Perimeter Institute for Theoretical Physics, 31 Caroline Street North, Waterloo, ON, N2L 2Y5, Canada}
\email{ptiede@perimeterinstitute.ca}
\author[0000-0002-3351-760X]{Avery E. Broderick}
\affiliation{Department of Physics and Astronomy, University of Waterloo, 200 University Avenue West, Waterloo, ON, N2L 3G1, Canada}
\affiliation{ Waterloo Centre for Astrophysics, University of Waterloo, Waterloo, ON N2L 3G1 Canada}
\affiliation{Perimeter Institute for Theoretical Physics, 31 Caroline Street North, Waterloo, ON, N2L 2Y5, Canada}
\email{abroderick@perimeterinstitute.ca}
\author[0000-0001-9625-5929]{Mohamad Shalaby}
\affiliation{Leibniz-Institut f{\"u}r Astrophysik Potsdam (AIP), An der Sternwarte 16, 14482 Potsdam, Germany}
\affiliation{Department of Physics and Astronomy, University of Waterloo, 200 University Avenue West, Waterloo, ON, N2L 3G1, Canada}
\affiliation{Perimeter Institute for Theoretical Physics, 31 Caroline Street North, Waterloo, ON, N2L 2Y5, Canada}
\affiliation{Department of Physics, Faculty of Science, Cairo University, Giza 12613, Egypt}
\email{mshalaby@perimeterinstitute.ca}

\author[0000-0002-7275-3998]{Christoph Pfrommer}
\affiliation{Leibniz-Institut f{\"u}r Astrophysik Potsdam (AIP), An der Sternwarte 16, 14482 Potsdam, Germany}
\email{cpfrommer@aip.de}
\author[0000-0001-8778-7587]{Ewald Puchwein}
\affiliation{Institute of Astronomy and Kavli Institute for Cosmology, University of Cambridge, Madingley Road, Cambridge, CB3 0HA, UK}
\affiliation{Leibniz-Institut f{\"u}r Astrophysik Potsdam (AIP), An der Sternwarte 16, 14482 Potsdam, Germany}
\email{epuchwein@aip.de}

\author[0000-0002-2137-2837]{Philip Chang}
\affiliation{Department of Physics, University of Wisconsin-Milwaukee, 3135 N. Maryland Ave., Milwaukee, WI 53211, USA}
\email{chang65@uwm.edu}
\author[0000-0001-8740-0127]{Astrid Lamberts}
\affiliation{Universit\'e C\^ote d'Azur, Observatoire de la C\^ote d'Azur, CNRS, Laboratoire Lagrange, Laboratoire ARTEMIS, France}
\email{astrid.lamberts@oca.eu}

\begin{abstract}
  Pair creation on the cosmic infrared background and subsequent inverse-Compton scattering on the CMB potentially reprocesses the TeV emission of blazars into faint GeV halos with structures sensitive to intergalactic magnetic fields (IGMF).  Previous work has shown that these halos are then highly-anisotropic and extended.  If the coherence length of the IGMF is greater than the inverse-Compton cooling length of the pairs, then the orientation of the gamma-ray halo will be correlated with the direction of the magnetic field which is unknown and expected to change for each source.  In order to constructively add each source we then use angular power spectra which are insensitive to the jet orientation. By looking at known GeV blazars detected by \Fermi, we exclude the existence of an IGMF with coherence lengths $>$100~Mpc at greater than $3.9\sigma$ with current-day strengths in the range $10^{-16}$ to $10^{-15}$~G, and at $2\sigma$ from $10^{-17}$ to $10^{-14}$~G. This provides a direct measurement of the non-existence of gamma-ray halos, providing an important check on previous results.
\end{abstract}


\section{Introduction}

Astronomical magnetic fields appear on scales ranging from the terrestrial to those of galaxy clusters, the latter spanning many Mpc.  Within all of these systems magnetic fields are believed to play a variety of important roles.  These include providing additional pressure components \citep{Beck2016}, mediating the magneto-hydrodynamic turbulence \citep{Goldreich_Sridhar1995}, providing and limiting the effective viscosity that transports angular momentum in accretion disks \citep{Balbus1991} and mediates collapse \citep{Hennebelle2012, Crutcher2012}, contributing to the acceleration of high energy particles, e.g., cosmic rays \citep{Blandford_Eichler1987}, and dictating their subsequent propagation \citep{Hanasz2003, Pakmor2016} as well as diffusion of thermal energy and momentum in weakly collisional and collisionless systems \citep{Schekochihin2009}.

Observed astrophysical magnetic fields are likely amplified from weak primordial seed fields.  These seed fields may originate from Biermann battery processes \citep{2008RPPh...71d6901K}, electroweak or quantum chromo-dynamics phase transitions in the early universe, or during cosmic inflation.  Biermann battery processes can naturally generate magnetic fields when temperature and density gradients are not parallel, resulting in electric fields that have a curl and providing a source for the magnetic field.  Typically, these processes have a correlation length that reflects the length scales associated with the temperature and density gradients \citep{2008RPPh...71d6901K,2012SSRv..166...37W}, and generate small scale ($\sim 10$ kpc) fields ($B \sim 10^{-20}$~G), which can be amplified by the action of a dynamo to $\mu$G fields observed in galaxies today \citep{Schober2013, Beck2016, Pakmor2017}.

By their nature, fields generated in primordial processes may be present in the intergalactic medium as the IGMF.  While the general nature of these fields in still unknown, causal constraints limit the coherent length scales of primordial fields generated from post-inflation mechanisms, e.g., phase transitions, to small comoving scales, typically $< 10$ kpc \citep{2012SSRv..166...37W}.    Small scale fields may seed large scale fields through the inverse cascade up to $\sim 0.1$ Mpc, i.e., the sonic scale of $10^4$ K gas over the age of the universe \citep{2012SSRv..166....1R}. 

To generate large-scale ($\gtrsim 1$ Mpc) volume-filling fields, seed fields must have been generated prior to or during inflation \citep{1950ZNatA...5...65B,1988PhRvD..37.2743T} or from second-order interactions between the photon and electron fluid prior to recombination \citep{2006Sci...311..827I}. In the former case, additional physics of super-adiabatic magnetic amplification \citep{2005PhRvD..71l3506T} is needed to keep these fields relevant today. This leaves the strength of these inflationary fields unconstrained \citep{2012SSRv..166...37W}.  In the latter case, the expected comoving field strength is $B\sim 10^{-24}$ G.

Probing these primordial fields is difficult, but some constraints can be set on a large-scale IGMF.  The best direct upper limits on the strength of a putative large-scale IGMF to date come from the impact of such fields on the Cosmic Microwave Background (CMB), roughly $10^{-9}$~G, followed closely by limits arising from Faraday rotation measurements, roughly $10^{-9}$ to $10^{-8}$~G depending weakly on the coherence length of the field $\lambda_B$ \citep{Nero-Vovk:10}.  Indirect limits on large-scale IGMFs can be obtained from the rotation measures of galaxy clusters, constraining the IGMF on scales exceeding a Mpc to be less than $10^{-12}$~G \citep{2005JCAP...01..009D,Nero-Vovk:10}.  In contrast, indirect lower-limits can be placed by the observed absence of inverse-Compton gamma-ray cascades.

Very-high energy gamma rays (VHEGRs), with energies exceeding $E>100$~GeV, lie above the pair-creation threshold with the infrared background \citep{Goul-Schr:67,Sala-Stec:98}.  As a result, the mean free path of VHEGRs in the nearby universe is $10^2 (1+z)^{-4.5} (E/6~\TeV)^{-1}$~Mpc \citep{PaperI}.  That is, while VHEGRs can traverse intergalactic distances, they are absorbed on cosmological distances, producing highly relativistic beams of electron-positron pairs (i.e., Lorentz factors of order $10^6$).  This redshift-dependent absorption of high energy gamma rays has been observed \citep{Fermi_EBL2012} and subsequently used to probe the infrared background \citep{Domi_etal:11,Gilm_etal:12,Domi_etal:13}.

In the absence of any other process, e.g., collective beam-plasma instabilities \citep{PaperI}, these beams will inverse-Compton cool by upscattering cosmic microwave background (CMB) photons to GeV energies.  That is, this chain of events would effectively reprocess the TeV emission of AGN into GeV emission.  Thus, bright, nearby, extragalactic TeV sources should exhibit a GeV bump that is correlated with their TeV spectral energy distribution (SED).  Stringent limits have been placed by \Fermi on the existence of such a GeV bump already, effectively excluding such a feature in several TeV sources \citep{Nero-Semi:09,Nero-Vovk:10}.

Presuming that inverse-Compton cooling dominates the evolution of the beams, these non-detections provide indirect, circumstantial evidence for the existence of a strong IGMF, with $B>10^{-15}$~G to $10^{-17}$~G, depending on the particular extragalactic background light model employed \citep{Nero-Vovk:10,Tayl-Vovk-Nero:11,Vovk+12}. This assumes that the duty cycle of the gamma-ray blazars is comparable to that observed for the radio mode in AGN, roughly $10^6$~yr; shorter duty cycles could result in GeV echos that lag behind the initial VHEGR emission by long times, resulting in a weaker limit: $B>10^{-17}$~G \citep{Derm_etal:10,Tayl-Vovk-Nero:11}.  Such an IGMF would deflect the resulting pairs, whose inverse-Compton emission would then be highly beamed away from us, naturally removing the GeV bump.  However, at the same time, this would open the door to the direct detection of the IGMF through the observation of GeV halos surrounding intrinsically gamma-ray bright objects, for which the inverse-Compton emission is beamed into our line of sight.

A number of searches for inverse-Compton halos using the publicly available \Fermi observations have been performed, yielding contradictory claims but no convincing evidence for their presence \citep{Ando+10,Nero-Semi-Tiny-Tkac:11,FLAT-stack:2013,Chen:2015}.  These have focused on the large radial extent of the gamma-ray halos, typically subtending many degrees.  However, this is generally degenerate with the point spread function (PSF) of \Fermi's Large Area Telescope (LAT), which until recently was highly uncertain \citep{Nero-Semi-Tiny-Tkac:11,LAT_perf}.  As result, even when detections are claimed they remain plagued by large systematic uncertainties.

  Previous work in \citet{Broderick+2018} stacked, rotated \Fermi images of Fanaroff-Riley class I and II objects from the VLA FIRST survey \citep{FIRST_CAT}, to limit the strength of the IGMF to $<10^{-15}G$ for large correlation length magnetic fields.  A key difference in this work is that we use \Fermi blazars and their full spectral information in the \GeV--\TeV\ range \citep[see][]{BowTiesII} to attempt to measure gamma-ray halos. This is different to the work by \cite{Broderick+2018} who statistically sampled the blazar SEDs using the distribution of \Fermi and VHEGR spectra. Due to this, the work in this paper provides an additional and independent constraint on the IGMF strength.

  The layout of this paper is as follows: In Section~\ref{sec:2}, we review our method first described in two companion publications~\citep{BowTiesI,BowTiesII}, where we used angular power spectra to coherently stack angular structure from gamma-ray halos. In Section~\ref{sec:3} we describe a model for the gamma-ray sky background, which adds power at low multipoles.  Section~\ref{sec:4}, details the main results followed by a discussion of the impact of various potential systematics.  Finally, Section~\ref{sec:5} summarizes the conclusions of this work.

\section{Method}\label{sec:2}
In two companion publications \citep{BowTiesI,BowTiesII} we have demonstrated that these inverse-Compton halos are expected to be highly anisotropic \citep[see also][]{2010ApJ...719L.130N,2015JCAP...09..065L}.  This originates from different physical origins for a small-scale, tangled IGMF ($\lambda_B\ll3$~Mpc) and a large-scale, uniform IGMF ($\lambda_B>100$~Mpc).  For the latter, of interest here, this is due to extreme beaming of the inverse-Compton emission coupled with the geometric conditions upon the gyration of the particles.  This typically produces a bilateral, narrowly beamed halo, the two lobes being due to the opposite gyration directions of the electrons and positrons \citep{BowTiesI}.

In \cite{BowTiesII} we demonstrated that existing \Fermi observations should permit the detection of a large-scale IGMF with strengths comparable to those implied by the absence of GeV bumps in nearby TeV sources, i.e., $10^{-16}$ to $10^{-15}$~G.  This makes use of stacked one-dimensional angular power spectra that amplify the impact of anisotropy\footnote{Note that \citet{Broderick+2018} used the VLA survey so that the jet direction is known, so stacking angular histograms coherently was possible.}: 
\begin{equation}
  {\mathcal P}_m = \frac{1}{N_{\rm src}} \sum_{\rm src}\frac{1}{N^2}\left|\sum_{j} e^{i m\theta_j}\right|^2\,,
\end{equation}
where $\theta_j$ is the polar angle of the $j$th photon defined relative to the position of the central source and an arbitrary direction, and $N$ the number of photons for the source.  For this we make use of the Pass 8R2\_V6 ULTRACLEANVETO event class and limit ourselves to gamma-ray energies between 1~GeV and 100~GeV to avoid large variations in the instrument response.  We further apply an energy-dependent and conversion-location-dependent mask (i.e., Front and Back), equal to the 68\% inclusion region of the Pass 8R2\_V6 PSF, thus excising the region most dominated by the direct emission from the source.  

The existence of a bilateral, jet-like feature appears primarily as an excess of quadrupolar power; with a sufficient number of photons it would produce a sawtooth-like structure in the power spectrum in which even multipoles exhibit an excess.  Based upon the source SEDs and Monte Carlo modeling of the \Fermi source population we have identified sets of \Fermi objects that are optimized for the detection of the anisotropic halo feature.  These necessarily depend upon the assumed strength of the IGMF, with larger IGMFs producing observable halos in more sources and therefore permitting a larger sample that may be profitably stacked.  These optimized source lists may be found in Table 1 of \cite{BowTiesII}, with the number of sources ranging from 4 to 18 for current-day strengths of $B_0=10^{-17}$~G to $10^{-14}$~G (note that the source-frame field strength is given by $B=B_0(1+z)^2$, though due to the small redshifts of the gamma-ray sources employed the two differ by at most a factor of 2).  For each, an uncontaminated region about each source has been identified via visual inspection.  Note that as a result these fields do typically contain many dim sources identified in the Fermi Large Area Telescope Third Source Catalog \citep[3FGL][]{3FGL}.

Critically, the anisotropy signal is unlikely to be confused with a number of anticipated systematic contaminants.  In principle, these may arise from unknown physics at the TeV source, the LAT response, or unmodeled structures in the gamma-ray sky.  In practice, chief among the potential contaminants is the degree-scale substructures in Galactic contribution to the diffuse gamma-ray background.  To assess the degree to which these contribute to the stacked angular power spectra we now construct a model of the gamma-ray background.


\begin{figure}
  \begin{center}
    \includegraphics[width=\columnwidth]{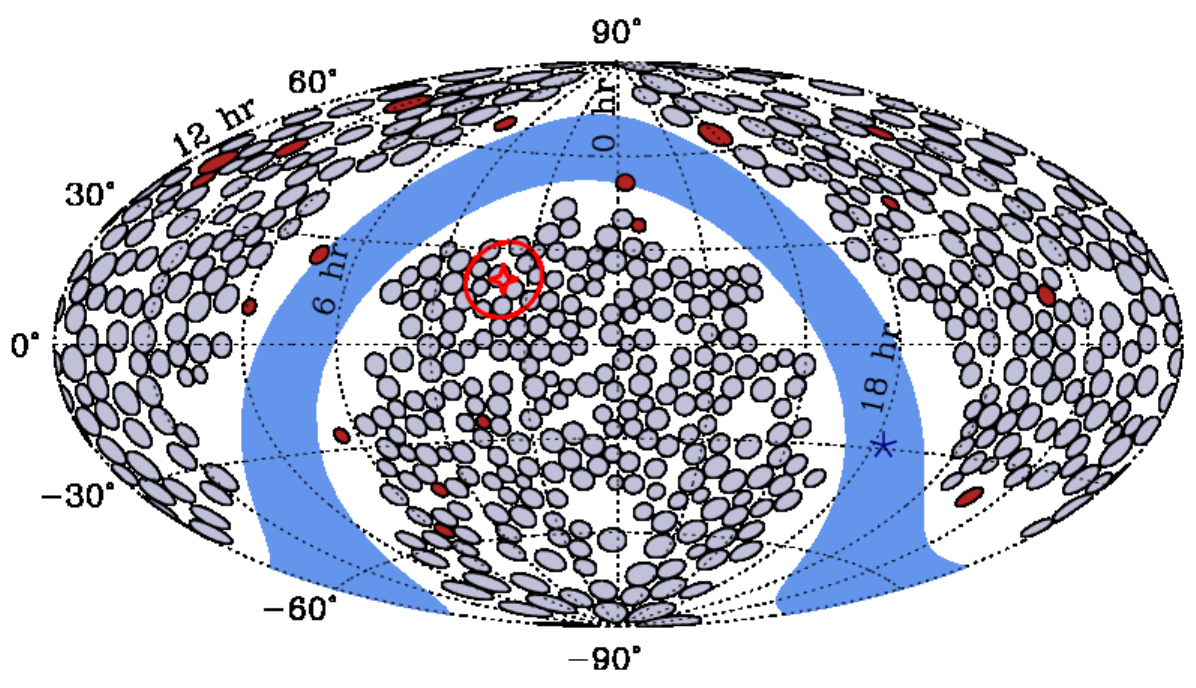}
  \end{center}
  \caption{Aitoff projection of the 18 source (red) and 507 high-latitude background (grey) field locations shown in equatorial coordinates.  The region with $|b|\le10^\circ$ is shown in blue and the location of the Galactic Center is shown by the asterisk.  For comparison, the location of 1ES~0229+200, the primary source employed by \cite{Tayl-Vovk-Nero:11}, is shown by the bright red star; the bright red circle indicates the projected 100~Mpc region about 1ES~0229+200.  All field locations are circles when viewed directly, the apparent asymmetries are due solely to projection.}\label{fig:locs}
\end{figure}

\section{Modeling the background contributions to the power spectra}\label{sec:3}

In this section we compute the contributions of the photons from background and contaminating 3FGL sources to the stacked angular power spectra.To assess, quantify, and guide the modeling of said contribution, we first identified 507 fields with Galactic latitudes more than $20^\circ$ from the Galactic plane.  The locations of these are shown in Fig.~\ref{fig:locs}.  Also shown for comparison are the source fields (red).  Note that while the background fields appear to nearly cover the sky, the bright point sources in the \Fermi sky have effectively been excised according to the criteria used in \cite{BowTiesII} to determine the size of the source fields.  That is, all of the bright sources fall into the gaps between the source and background fields.  Note that these fields are not devoid of 3FGL sources -- sources with fluences less than roughly 40~photons do appear in these background fields and in the source fields.  The stacked power spectrum of these 507 fields comprise the background power spectrum that we sought to successfully model.

The model we constructed consists of four components:
\begin{enumerate}
\item 3FGL sources
\item Completion of the 3FGL below the survey sensitivity limit
\item Large-scale gradients ($>9^\circ$)
\item Small-scale gradients ($\approx2^\circ$)
\end{enumerate}
We will discuss the construction of each individually.
\begin{figure*}
  \begin{center}
    \includegraphics[width=0.48\textwidth]{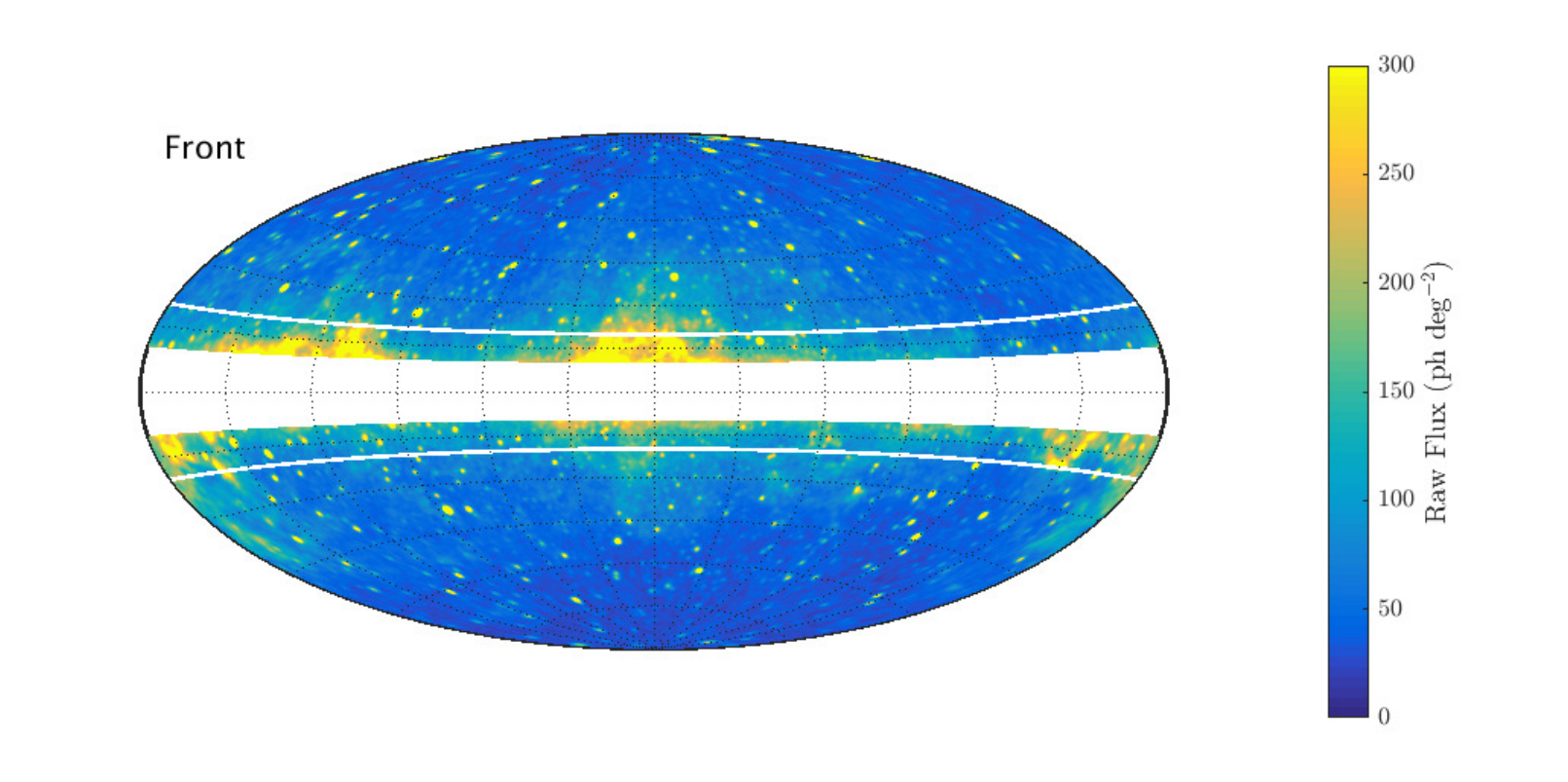}
    \includegraphics[width=0.48\textwidth]{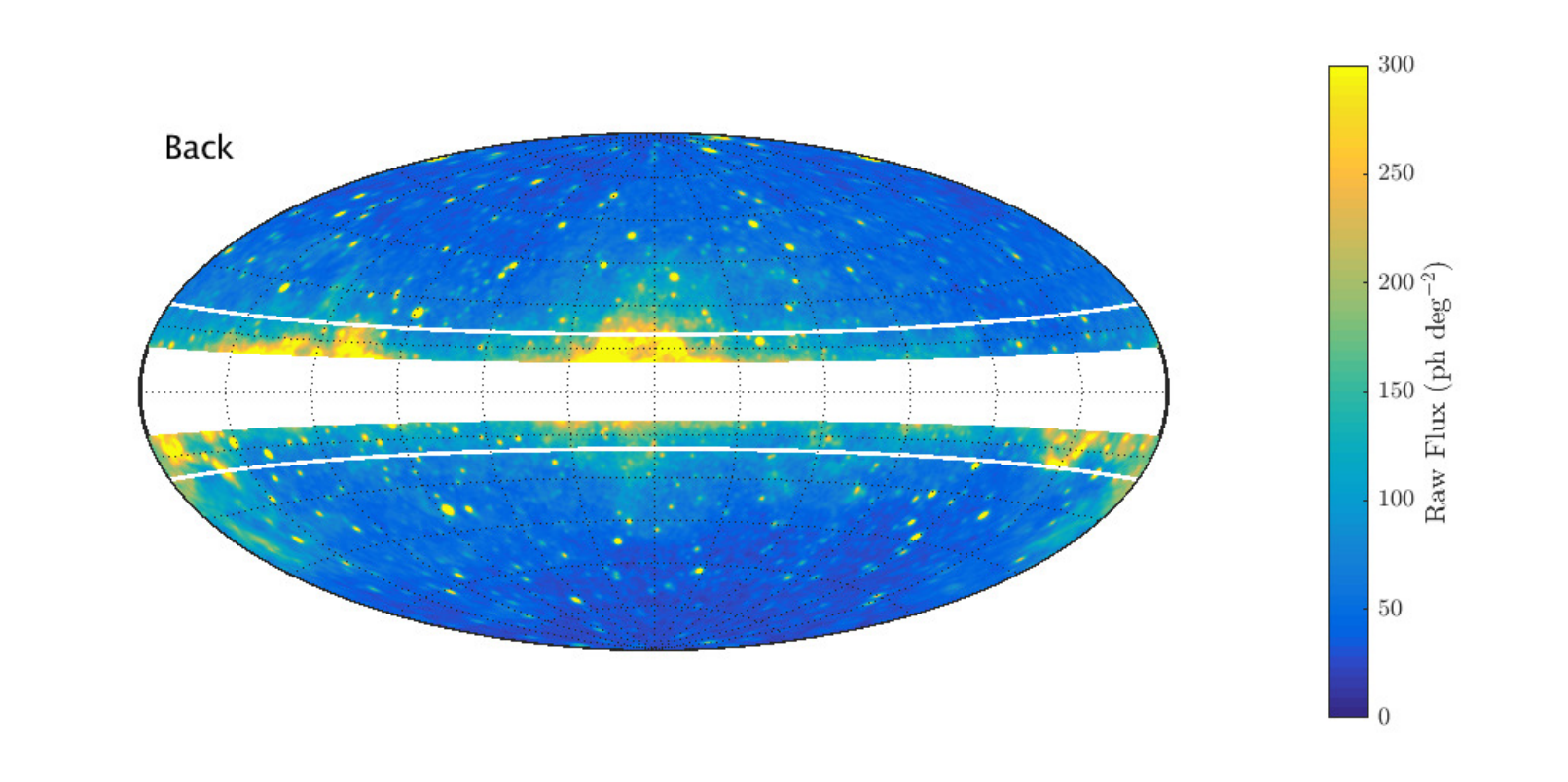}\\
    \includegraphics[width=0.48\textwidth]{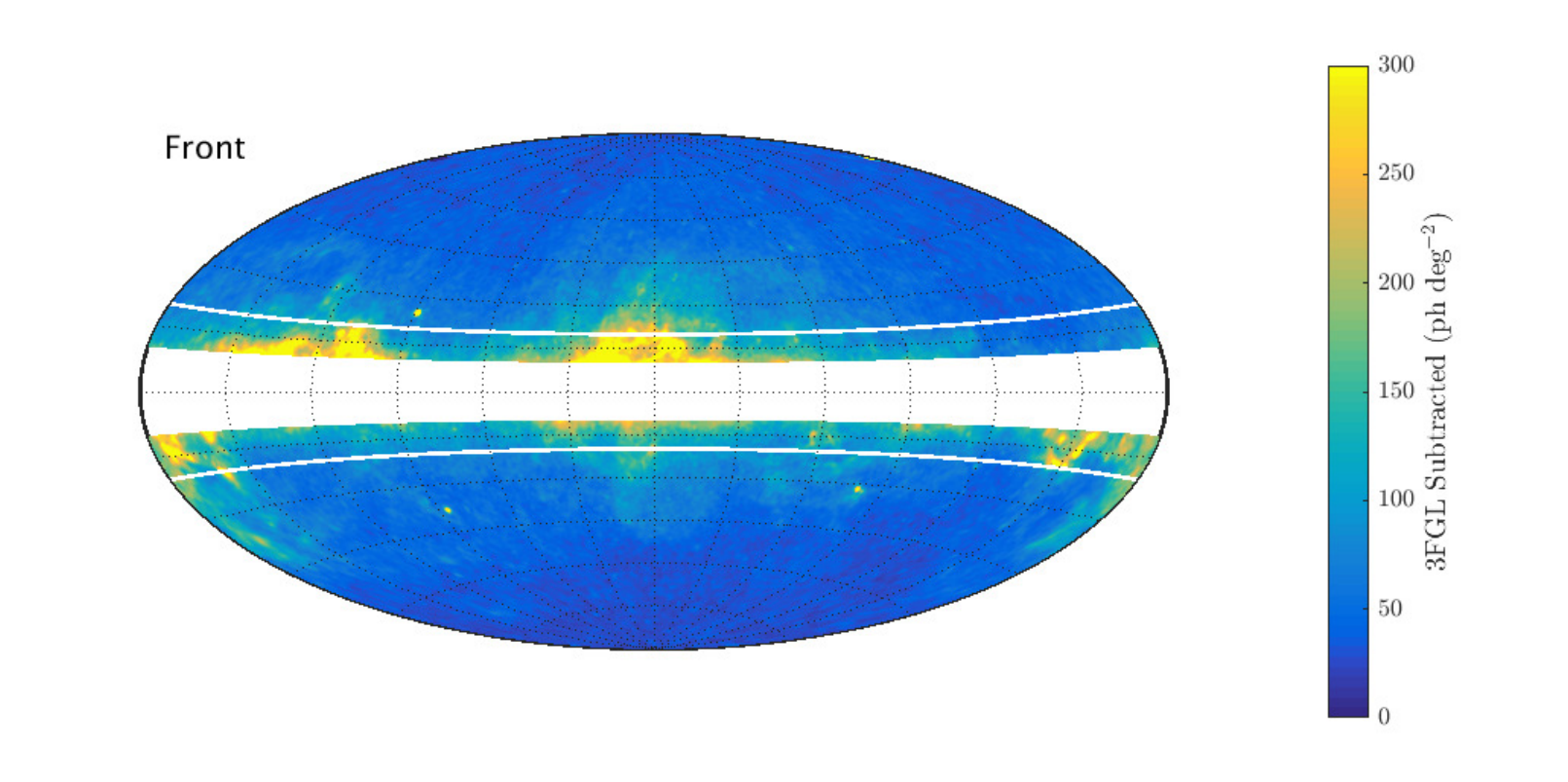}
    \includegraphics[width=0.48\textwidth]{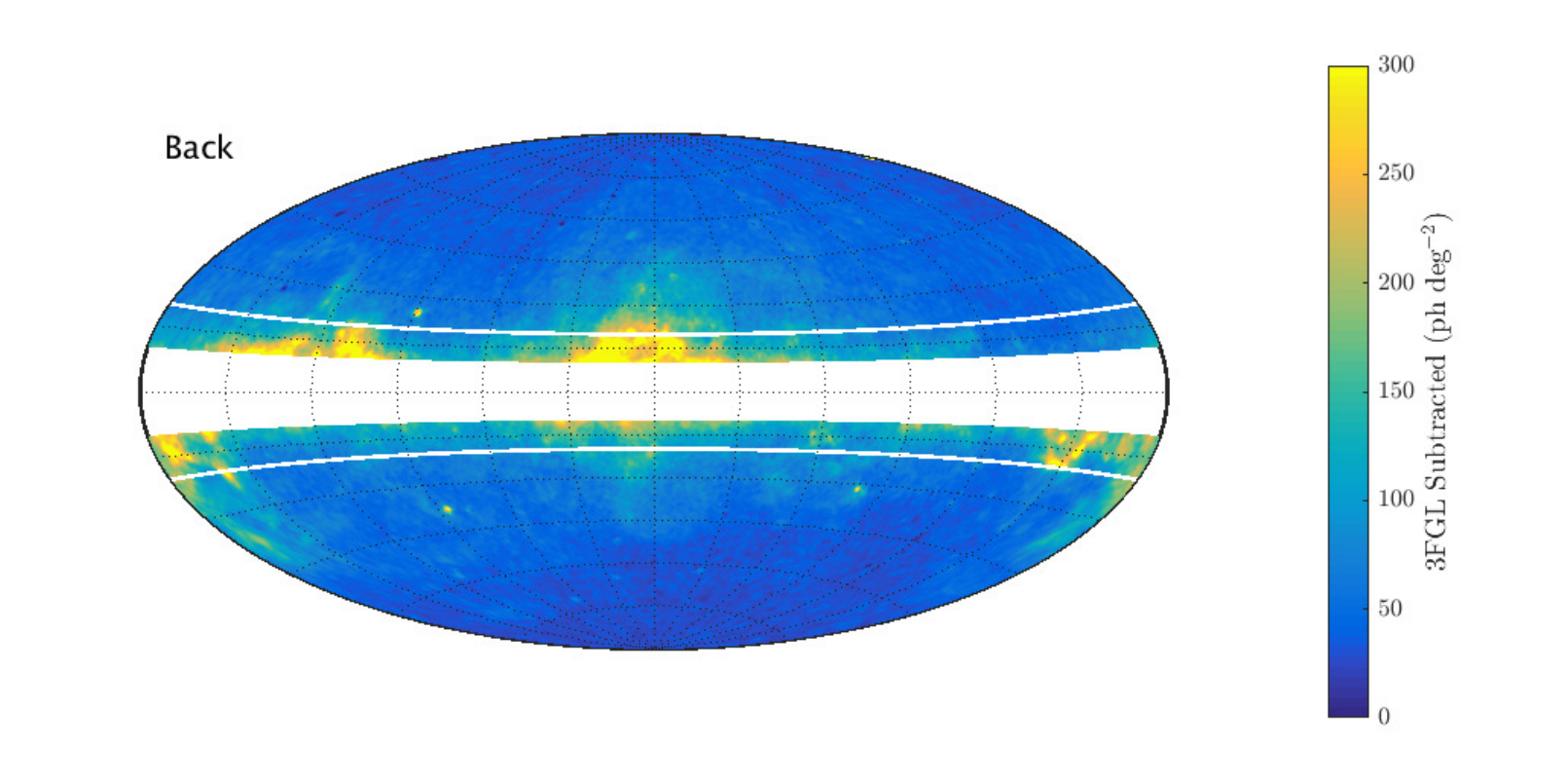}\\
    \includegraphics[width=0.48\textwidth]{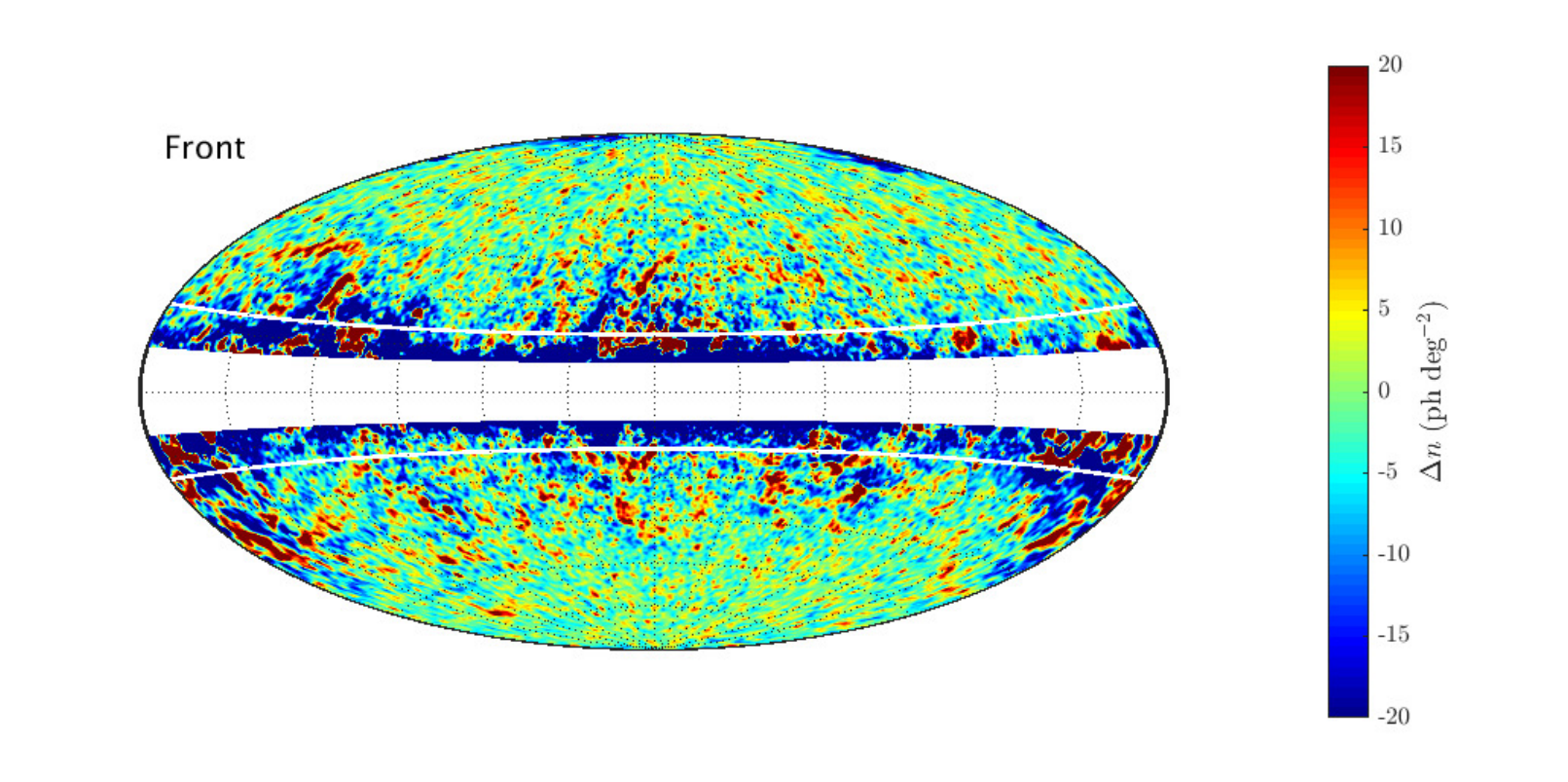}
    \includegraphics[width=0.48\textwidth]{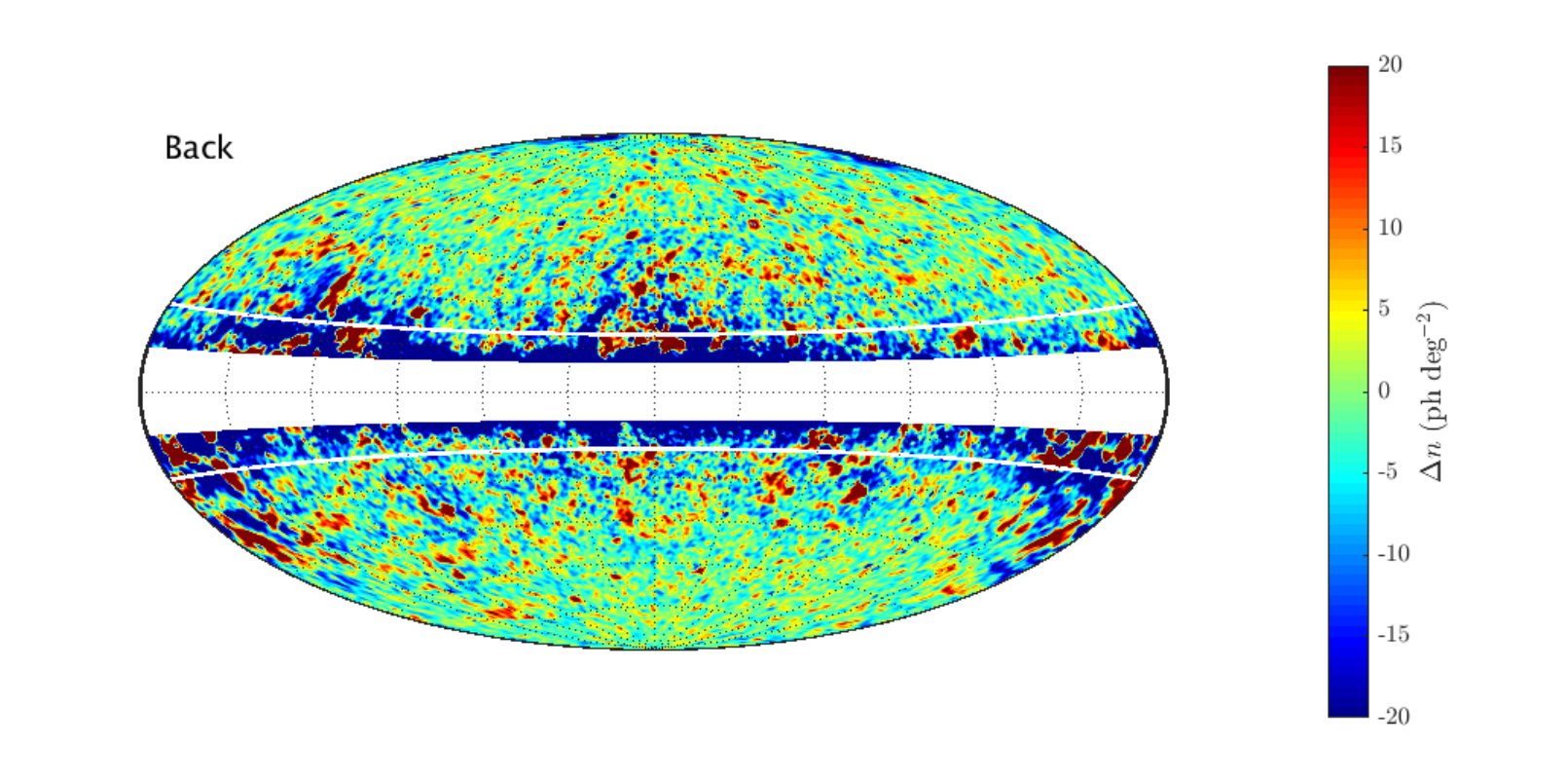}\\
    \includegraphics[width=0.48\textwidth]{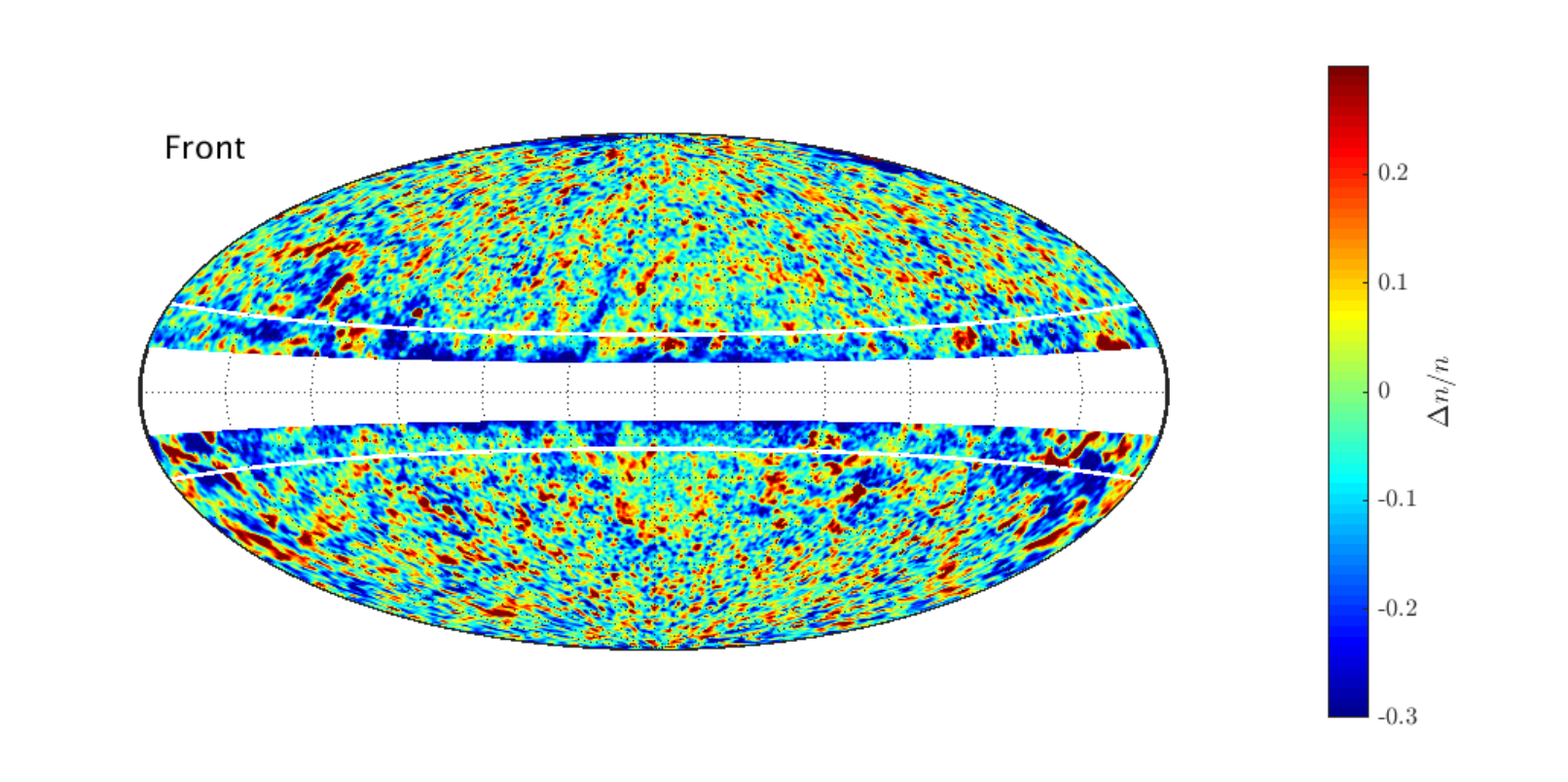}
    \includegraphics[width=0.48\textwidth]{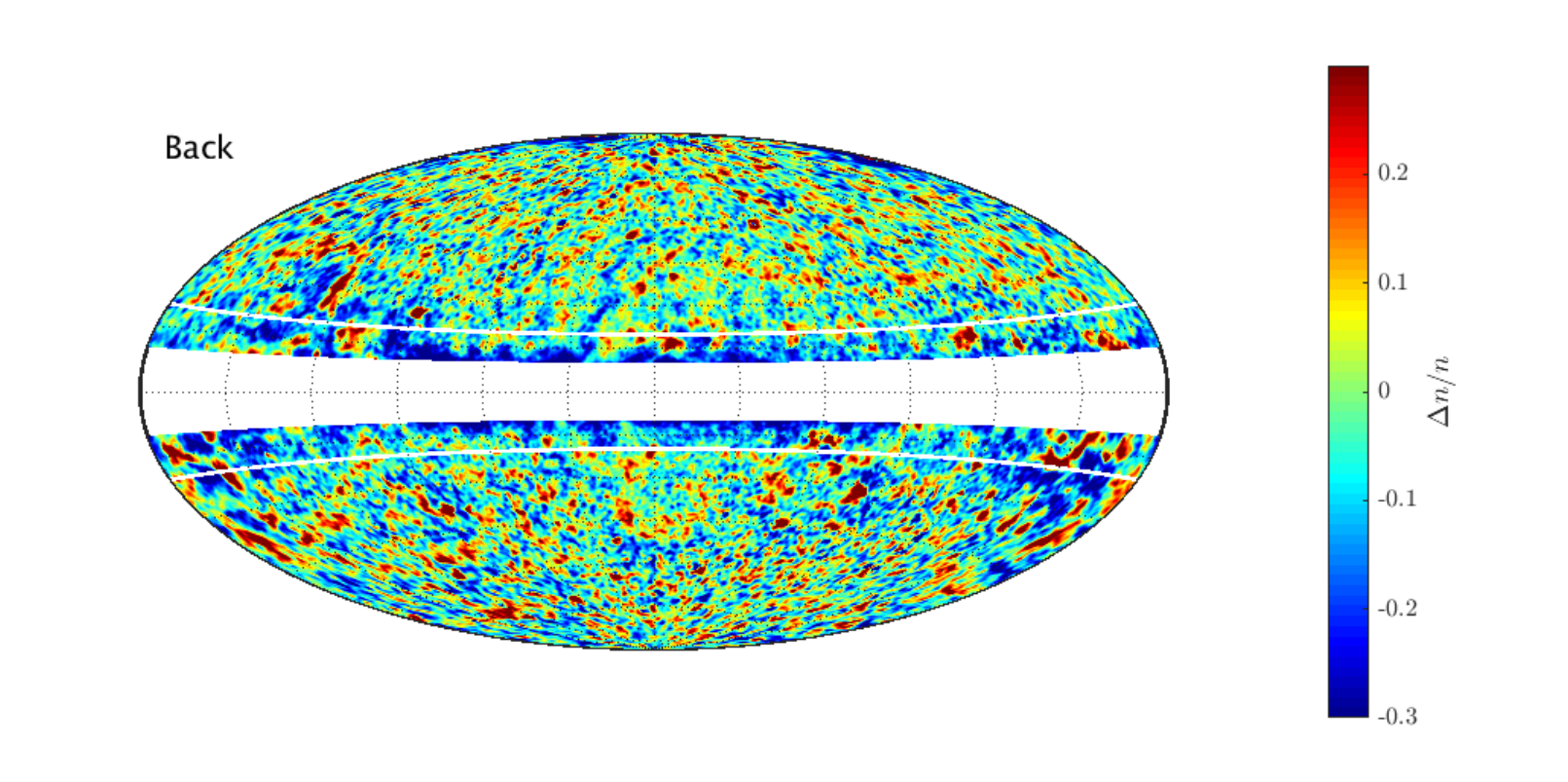}\\
  \end{center}
  \caption{Aitoff projected maps in Galactic coordinates of the gamma-ray sky at 1-100~GeV for events converted in the front (left) and back (right) detectors.  Top: the flux density smoothed with a Gaussian filter with a FWHM of $1.2^\circ$.  Upper-middle: the flux density smoothed with a Gaussian filter with a FWHM of $1.2^\circ$ after subtracting the 3FGL sources with their refitted fluences.  Lower-middle: the difference of 3FGL subtracted maps smoothed over $1.2^\circ$ and $9.4^\circ$, where the latter is a measure of the large-scale structure in the map, producing an image of the small-scale structure after removal of the 3FGL.  Bottom: the fractional difference of the 3FGL subtracted maps smoothed over $1.2^\circ$ and $9.4^\circ$, where the latter is taken as the normalization.  In all plots lines of $|b|=20^\circ$ are shown in white, and the region for which $|b|\le10^\circ$ has been masked out to remove the Galactic plane.}\label{fig:maps}
\end{figure*}

\subsection{Inspecting the \Fermi sky}
The construction of all of these is facilitated by the direct inspection of the \Fermi sky, obtained from 239557417~s (the official start of the \Fermi data release) to 487012885~s in mission elapsed time (MET). A density map of the 1-100~GeV photons in the Front- and Back-converted Pass 8R2\_V6 ULTRACLEANVETO events class is smoothed with a Gaussian kernel.  The size of this kernel is informed by typical structure of the point sources, set by the Pass 8R2\_V6 PSF, below which no discernible structure appears.  Thus, as our finest resolution map we choose a kernel full-width, half-max (FWHM) of $1.2^\circ$, which is shown in the top row of Fig.~\ref{fig:maps}.  The multitude of point sources that comprise the 3FGL are immediately evident.

\subsection{Component 1: 3FGL Sources}
Because the period from which the 3FGL was constructed is roughly half that over which these maps are generated, both exposure differences and source variability will produce significant differences in the anticipated number of photons from each 3FGL object.  Therefore, we begin by refitting the individual 3FGL source photon fluences.

We do this by comparing the observed gamma-ray maps smoothed on $1.2^\circ$ and $4.7^\circ$ scales to the expected response from the individual point sources.  That is, we begin by computing the observed photon density at each of the 3FGL source locations, given by $n^{1.2}_j$ and $n^{4.7}_j$, respectively, for the $j$th source.  We then simulate the corresponding anticipated fractional response of each individual source at each of the 3FGL positions, i.e., the number density of photons located at the $j$th source location associated with $N_i$ photons from the $i$th source are $f_{ji}^{1.2} N_i$ and $f_{ji}^{4.7} N_i$ for the two smoothing scales, respectively.  In doing this we employ the Pass 8R2\_V6 ULTRACLEANVETO PSFs and assume the spectral index listed in the 3FGL.

In addition to the point sources in the 3FGL there exists a diffuse component associated with gamma-ray emission from Galactic leptons and hadrons.  Unlike point sources, this contribution should be similar at both smoothing scales.  Thus, we have
\begin{equation}
  n^s_j = \bar{n}_j + \sum_i f^s_{ji} N_i
\end{equation}
for some fixed set of diffuse background photon densities, $\bar{n}_j$, at both smoothing scales.  This may be concurrently solved for the $\bar{n}_j$ and $N_i$, yielding the set of linear equations for the latter,
\begin{equation}
  \sum_i \left(f^{1.2}_{ji}-f^{4.7}_{ji}\right) N_i = \left(n^{1.2}_j - n^{4.7}_j\right)\,.
\end{equation}
We solve this for the $N_i$, and subsequently verify that the implied photon density associated with the diffuse background is positive.  While this method for estimating the 3FGL source fluences does account for the differing exposures and source variability, it does not account for the possibility of new sources.  It is also poorly suited to correcting the fluences of extended sources.  We do not anticipate either of these limits to present significant obstacles.

With the individual source fluences we then construct a realization of the 3FGL, smooth it over the appropriate scale, and subtract it from the observed photon density.  This is shown for a smooth scale of $1.2^\circ$ for Front- and Back-converted Pass 8R2\_V6 ULTRACLEANVETO events in the second row of Fig.~\ref{fig:maps}.  Apart from a handful of extended Galactic sources, and up to Poisson fluctuations in the number of sources for the Back-converted map, point sources have been reasonably well removed.  In practice, because the background fields specifically exclude bright sources, the details of the 3FGL removal are relatively unimportant.  Simply subtracting the 3FGL sources assuming the 3FGL fluxes and fixed exposure produces qualitatively identical results.

\subsection{Component 2: Completion of the 3FGL}
\begin{figure}
  \begin{center}
    \includegraphics[width=\columnwidth]{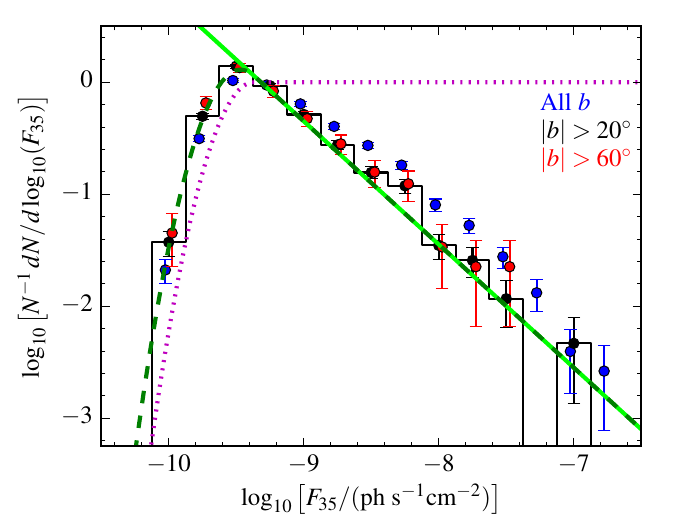}
  \end{center}
  \caption{Flux distribution of 3FGL sources with Galactic latitudes above $20^\circ$ (black).  For comparison the flux distribution of 3FGL sources without any Galactic latitude cut (blue points) and an extreme latitude cut ($|b|>60^\circ$, red points) are also shown; these are offset for clarity.  In green we show the single power-law model we assume for the intrinsic flux distribution.  The purple dotted line shows the assumed detection efficiency, $w(F_{35})$, and the dark-green dashed line shows the associated expected flux distribution, both evaluated at $b=90^\circ$.}\label{fig:3fglFdist}
\end{figure}
The second component, the unresolved completion of the 3FGL, can only be inserted in a statistical sense.  Ultimately, a complement of sources with the appropriate fluxes are selected and randomly placed within the background/source fields in a manner consistent with the $9.4^\circ$-smoothed background.  The source fluxes are chosen from the power-law distribution shown in Fig.~\ref{fig:3fglFdist}:
\begin{equation}
N^{-1} \frac{d N}{d\log_{10} F_{35}} = 0.45 \left(\frac{F_{35}}{10^{-9}~{\rm ph~s^{-1} cm^{-2}}}\right)^{-1.1}.
\end{equation}
This is supplemented with a detection efficiency, which we assume has a log-normal cutoff, i.e.,
\begin{equation}
  w(F_{35}) = \begin{cases}
    \exp[-12.88 (\log_{10}F_{35}/F_{\rm th})^2] & F_{35}<F_{\rm th}\\
    1 & \text{otherwise.}
  \end{cases}
\end{equation}
\begin{figure}
  \begin{center}
    \includegraphics[width=\columnwidth]{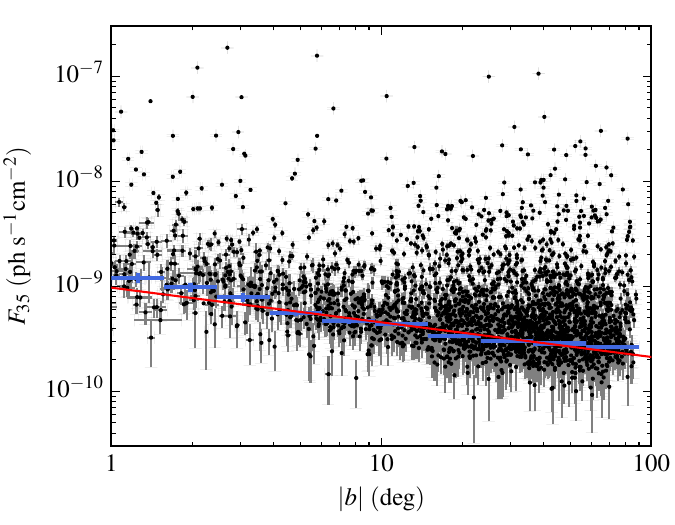}
  \end{center}
  \caption{Flux versus Galactic latitude for the 3FGL catalog. Blue points indicate the estimated position of the lowest-flux quartile in the indicated latitude bin; error estimates are obtained via bootstrap and incorporate the underlying flux uncertainties. The threshold for detection evolves with latitude as $b^{-0.33}$, shown for an appropriate normalization by the red line.} \label{fig:Fth}
\end{figure}

While for the $|b|>20^\circ$ population the threshold flux, $F_{th}$, is well described by a single number, in detail it does depend on Galactic latitude.  This is clear in Fig.~\ref{fig:Fth}, in which there is a trend in the lower envelope of 3FGL fluxes with Galactic latitude, scaling roughly as $|b|^{-0.33}$.  Therefore, we set
\begin{equation}
  F_{\rm th} = 4.27\times10^{-10} |b/90^\circ|^{-0.33} {\rm ~ph\,cm^{-2}\,s^{-1}}\,.
\end{equation}

Note that, despite the fact that the assumed flux distribution produces a formally infinite flux (and is thus unphysical), the condition that photons are observed in integer numbers, i.e., at least one, limits the number of photons observed in practice.  That is, for a flux $F_{35}$ the probability that at least a single photon will be observed is
\begin{equation}
  p_1(F_{35}) = 1 - p_0(F_{35}) = 1-e^{-F_{35} \epsilon} \approx F_{35} \epsilon\,,
\end{equation}
where $\epsilon$ is the exposure.  Thus, at small fluxes, the probability of observing at least one photon is linearly suppressed.  Therefore, any flux distribution flatter than $F_{35}^{-2}$ produces a convergent number of photons.

In practice, the contribution of the unresolved completion of the 3FGL is negligible.
\begin{figure}[!th]
  \begin{center}
    \includegraphics[width=\columnwidth]{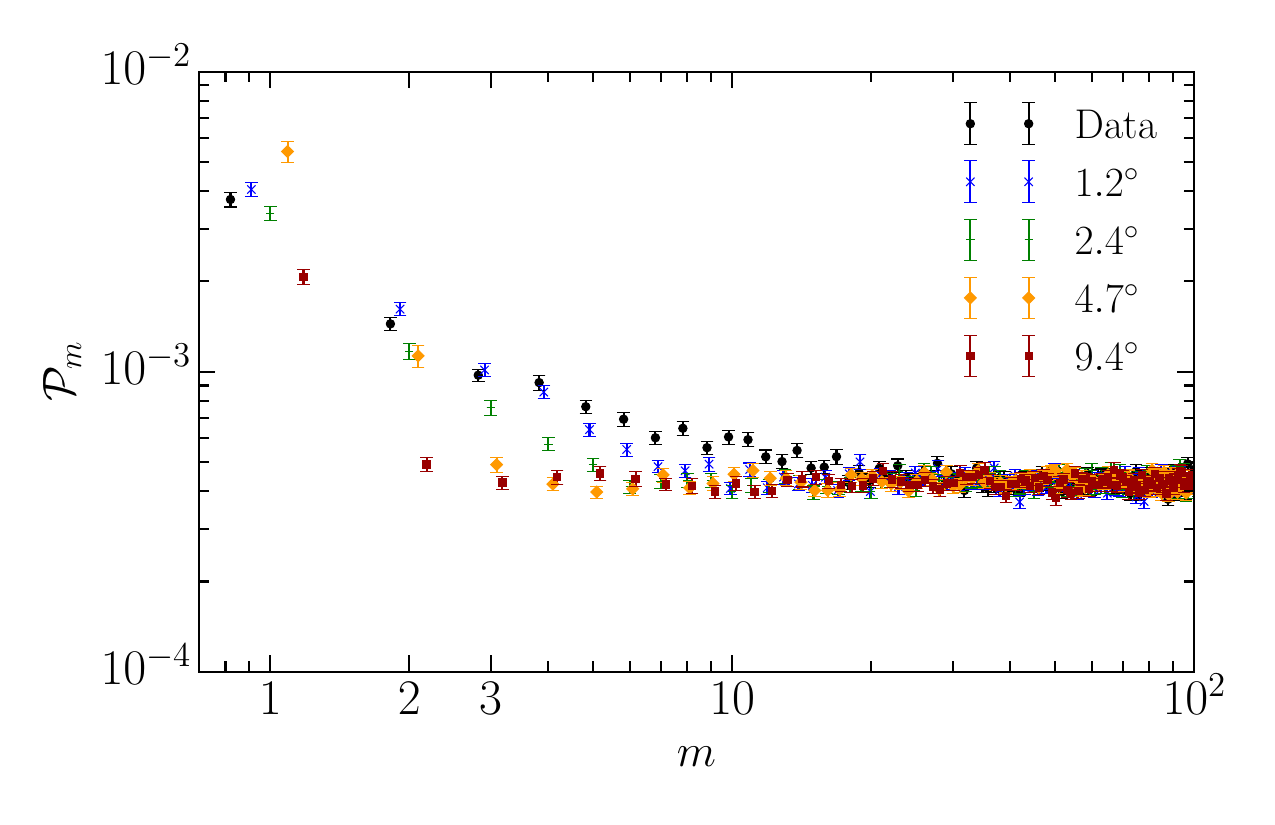}
  \end{center}
  \caption{Stacked angular power spectra of the background fields from the data (black), and for realizations drawn from Gaussian smoothed background maps with various smoothing scales.  As the smoothing scale increases power is suppressed at progressively smaller $m$.  Power at $m\le2$ is associated with structures on roughly a few degree scales.}\label{fig:smooth}
\end{figure}
\subsection{Component 3 and 4: Gradient fields}
The third component consists of randomly drawing the remaining background photons with a spatial distribution that follows the $9.4^\circ$-smoothed map.  This modifies the dipole component significantly, though again makes a negligible difference at $m>1$ as shown in Fig.~\ref{fig:smooth}.
\begin{figure}[!th]
  \begin{center}
    \includegraphics[width=0.48\textwidth]{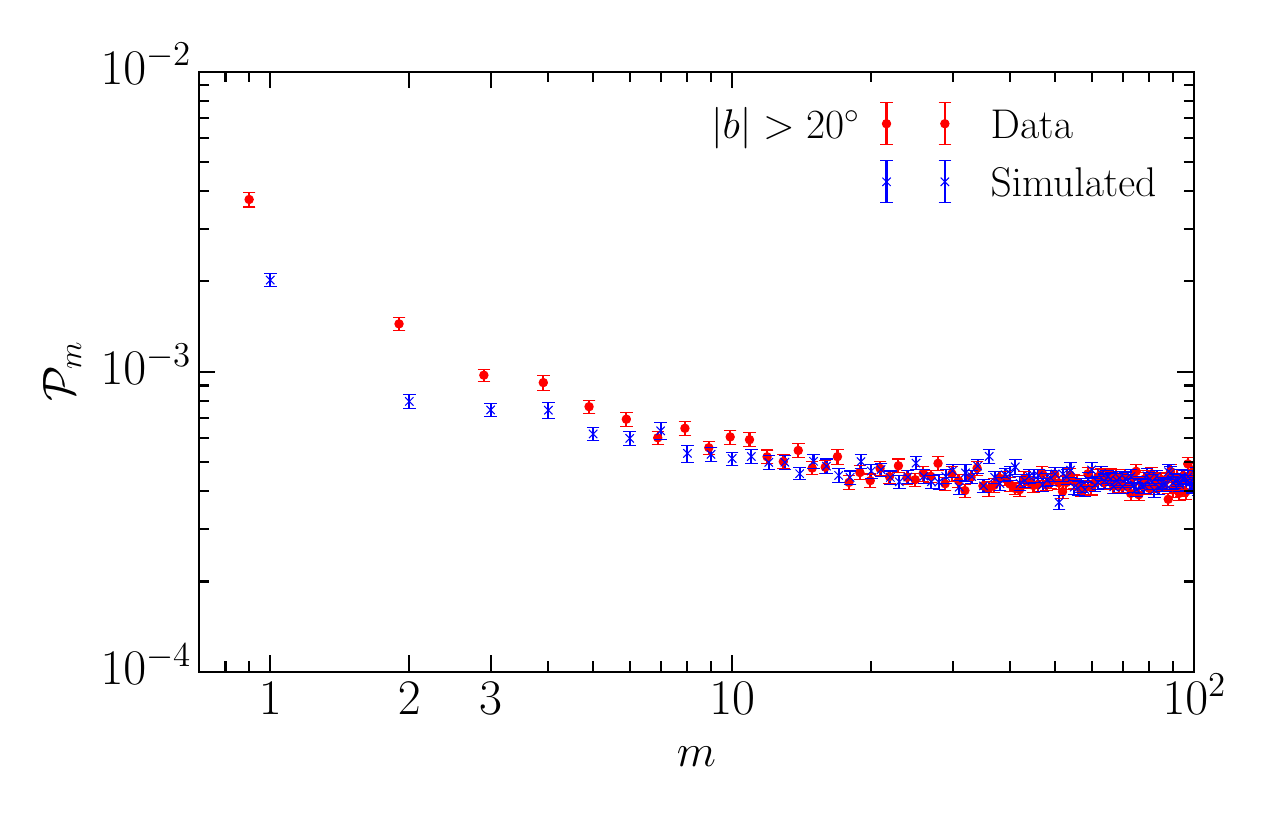}
    \includegraphics[width=0.48\textwidth]{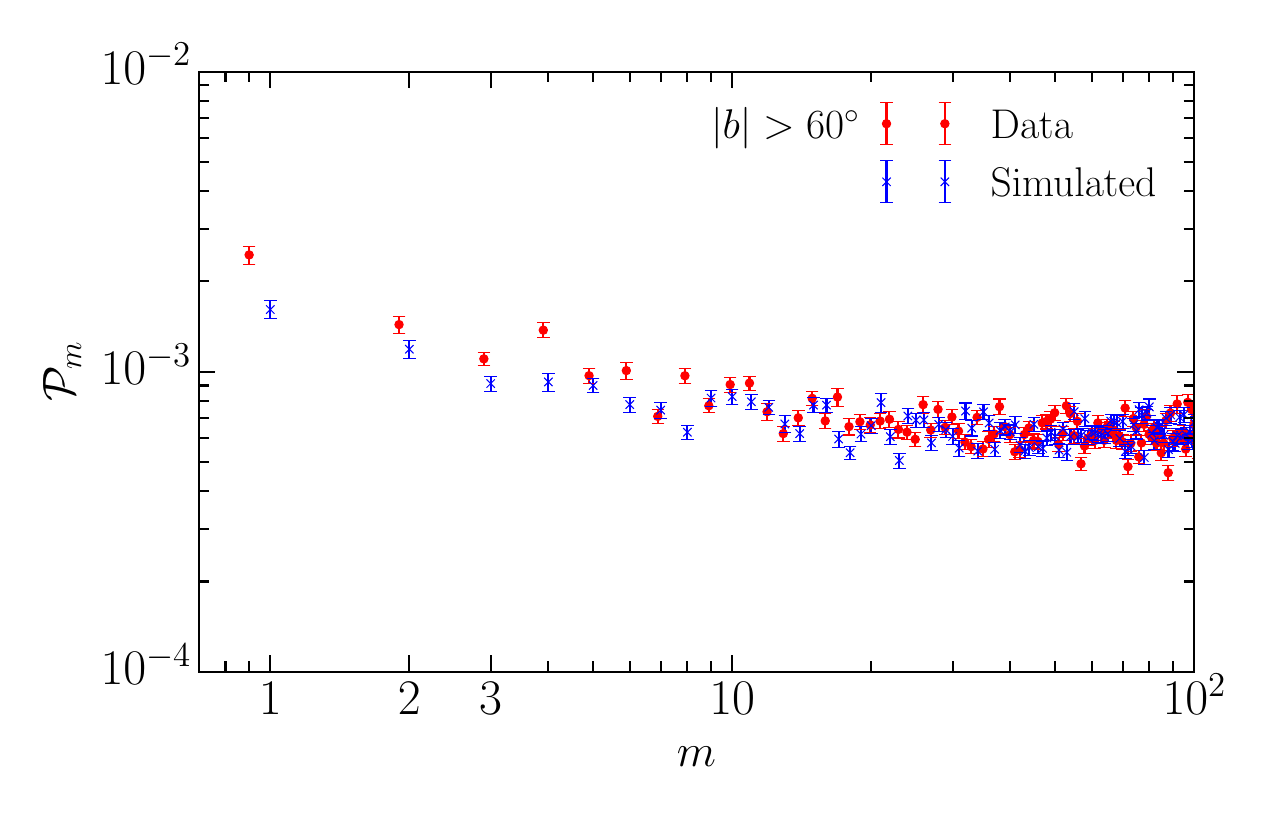}
  \end{center}
  \caption{Stacked angular power spectra for background fields above a given Galactic latitude from the background fields from the data (red), and for simulated realizations that include the 3FGL, an approximation of its completion, and the diffuse structure on scales above $9.4^\circ$ (blue).  The comparison becomes increasingly good as the Galactic latitude increases, strongly suggesting that the disparity is a result of structure in the Galactic foreground.} \label{fig:Plat}
\end{figure}

Finally, the fourth component consists of an attempt to model the degree-scale fluctuations in the diffuse Galactic background.  Stacked angular power spectra generated from realizations of the background fields using smoothed maps of the gamma-ray sky are shown in Fig.~\ref{fig:smooth}.  Unsurprisingly, larger smoothing scales correspond to less power at large $m$.  Specifically, at smoothing scales of $1.2^\circ$ the power above $m=5$ is noticeably suppressed.  Beyond smoothing scales of $2.4^\circ$ this suppression extends to $m\ge3$; by $9.4^\circ$ only the dipole power remains.  This suggests that any missing power at $m=2$ is most sensitive to a few degree-scale structures in the diffuse gamma-ray background.  Therefore, to clearly identify these we generate difference maps between high ($1.2^\circ$) and low ($9.4^\circ$) smoothing scales.  These are shown in the third row of Fig.~\ref{fig:maps}.

\begin{figure}
  \begin{center}
    \includegraphics[width=\columnwidth]{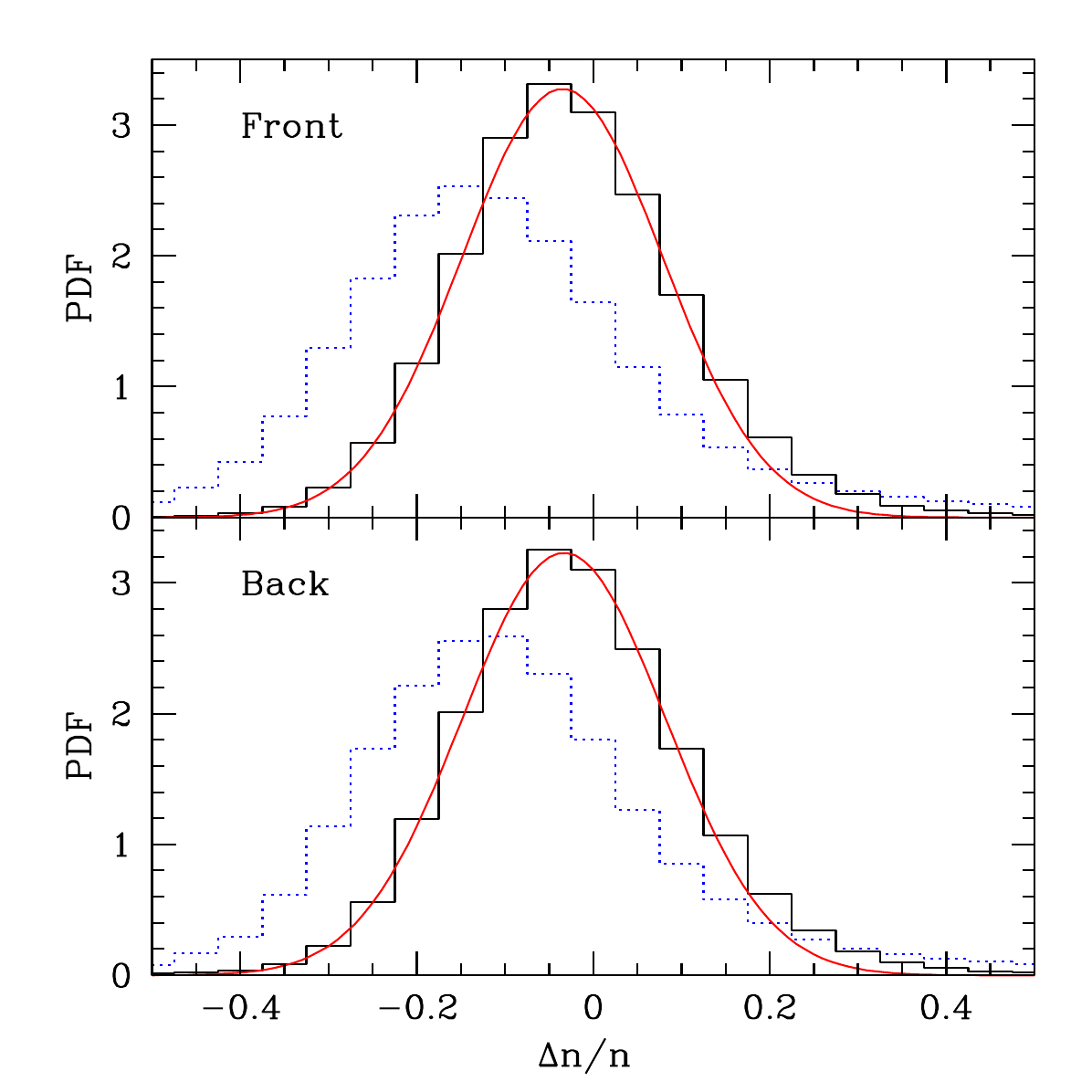}
  \end{center}
  \caption{Fractional flux probability distribution function for fractional fluctuations in the 3FGL subtracted maps of Front-converted (top) and Back-converted (bottom) events, smoothed over $1.2^\circ$ and $9.4^\circ$.  A Gaussian fit is shown in red for each, with mean -0.035 and -0.033, standard deviation 0.114 and 0.115, for the Front- and Back-converted event distributions, respectively.  In both cases the relative norm of the Gaussian is 0.94.  For comparison, in the blue dotted line is the PDF of fractional fluctuations obtained without subtracting the 3FGL sources, giving an indication of the impact of the unresolved completion of the 3FGL.  That is, the negative mean and substantial tails at large deviations are likely to be associated with the incomplete source subtraction.} \label{fig:fPDF}
\end{figure}

\begin{figure}
  \begin{center}
    \includegraphics[width=\columnwidth]{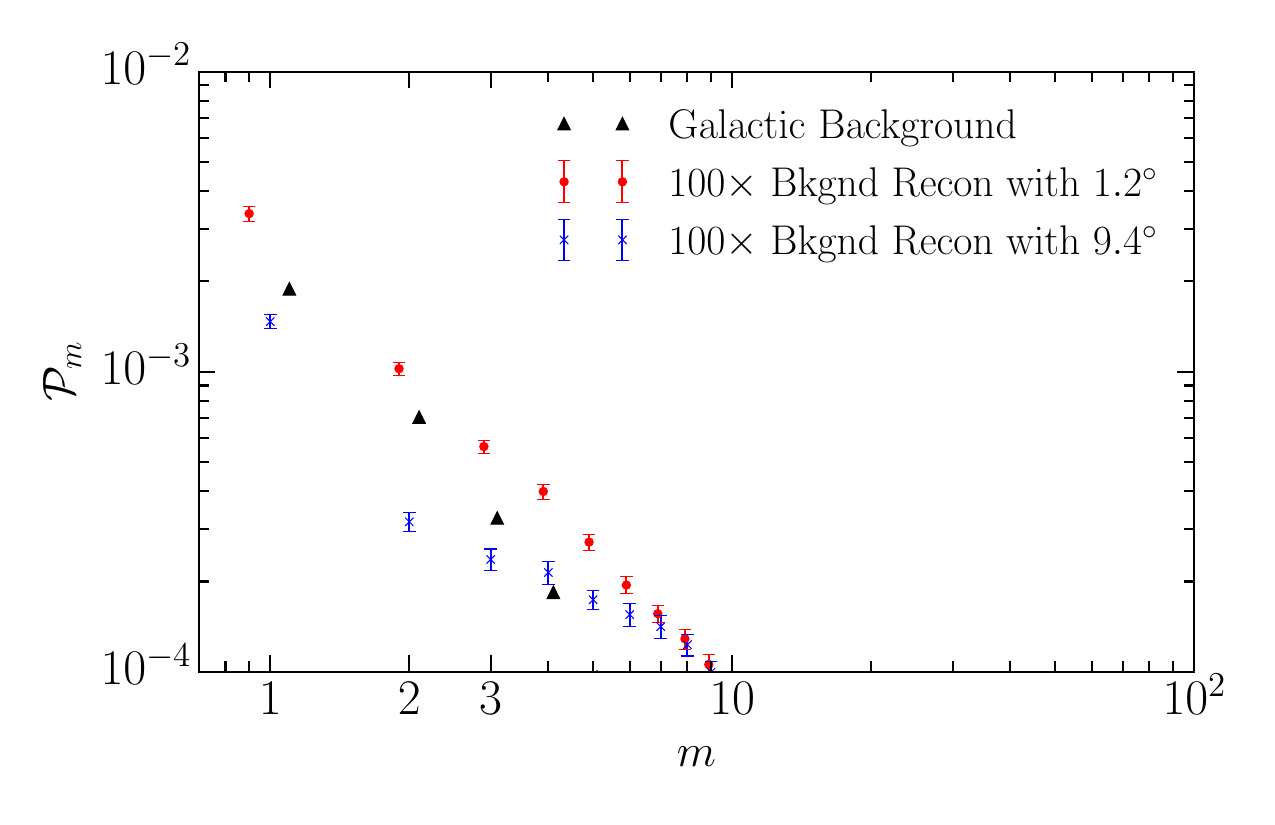}
  \end{center}
  \caption{Comparison of the stacked angular power spectra for simulated background fields using the smoothed, 3FGL-subtracted sky maps smoothed on $1.2^\circ$ and $9.4^\circ$ to reconstruct the degree-scale Galactic contribution.  To prevent the Poisson noise from dominating the estimate, we randomly draw 100 times more photons from the flux maps than present in the background fields.  We model this component as a multipole-dependent log-normal additional component with means and standard deviations chosen to reproduce the difference in power and difference in the variances of the low- and high-resolution angular power spectra. The black triangles show the angular power spectra of the Galactic background of the 3FGL source subtracted \Fermi sky.} \label{fig:ures}
\end{figure}

Generally, these exhibit a latitude dependent amplitude, falling rapidly with distance from the Galactic plane.  This supports their association with Galactic structures.  A similar conclusion holds regarding their impact on the stacked angular power spectra.  Figure \ref{fig:Plat} shows a direct comparison between the observed and simulated stacked power spectra, including only the first three components, from the background fields assuming different latitude cuts.  When restricted to fields with $|b|>60^\circ$ the two are very similar, as anticipated by the large reduction of the Galactic diffuse emission at high latitudes.

In contrast to their absolute amplitudes, the fractional amplitudes in the photon densities are roughly fixed, however (shown in the bottom row of Fig.~\ref{fig:maps}).  The PDF of the fractional density fluctuations is well described by the normal distributions, shown in Fig.~\ref{fig:fPDF}.  The mean is offset significantly from zero.  This is a natural result from imperfect point source subtraction, likely arising from the failure to subtract unidentified sources.  This is supported by the PDF of fractional density fluctuations constructed without any point source subtraction, shown in the dotted blue lines in Fig.~\ref{fig:fPDF}, which also shows a noted shift to negative fractional flux fluctuations.

The impact of the degree-scale structures in the Galactic diffuse emission is to increase the power at low $m$.  We quantify this by generating two stacked angular power spectra for the background fields from over-resolved realizations of the gamma-ray sky that include only the initial three components: 3FGL, its completion, and large-scale gradients (Figure~\ref{fig:ures}).  These differ in the maps used to generate the last component, the large-scale gradients; one using the map smoothed on $9.4^\circ$ and the other smoothed on $1.2^\circ$.  The latter map is statistically indistinguishable from the stacked angular power spectrum obtained from the data directly.  The difference between the two is entirely due to the otherwise unresolved degree-scale structures of interest.

\begin{figure}[th!]
  \begin{center}
    \includegraphics[width=0.48\textwidth]{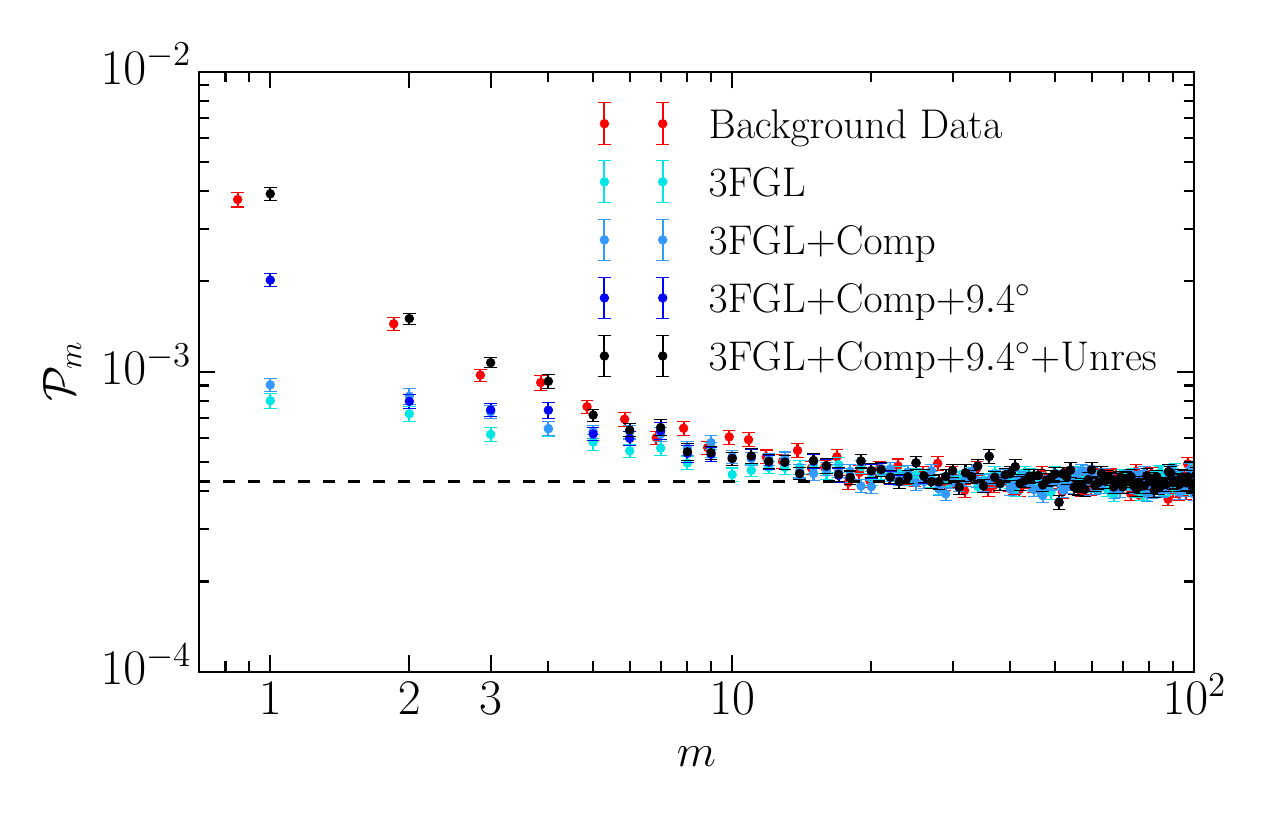}
  \end{center}
  \caption{Comparison of the observed (red) and reconstructed (black) stacked angular power spectra for the background fields.  The formal 1$\sigma$ errors for all 507 background fields are shown.  The reconstructed power spectra are constructed using simulated realizations that include the 3FGL, an approximation of its completion, the diffuse structure on scales above $9.4^\circ$, and the degree-scale background fluctuations.}\label{fig:bkgd_sum}
\end{figure}

\begin{figure}[!ht]
  \begin{center}
    \includegraphics[width=0.48\textwidth]{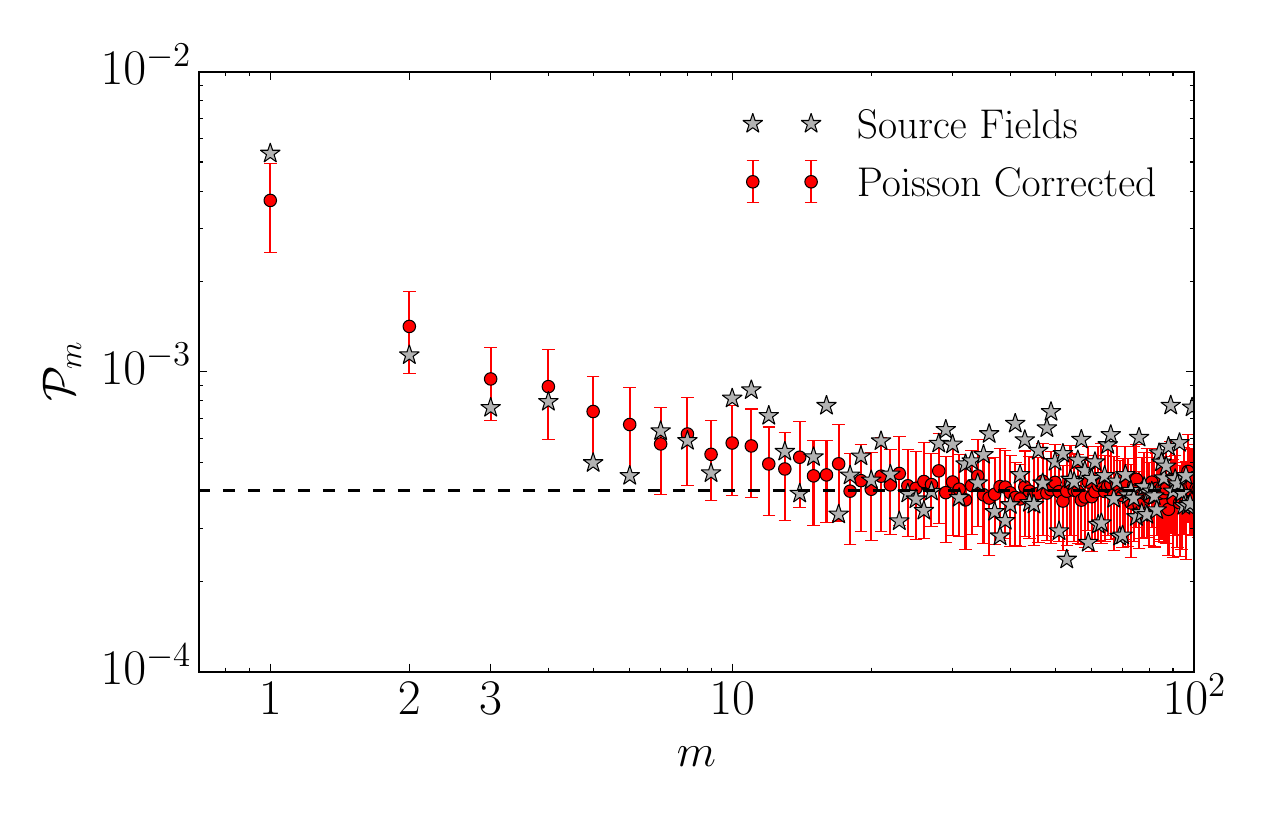}
  \end{center}
  \caption{Comparison of the stacked angular power spectra for the source (optimized for $10^{-15}$~G) and 507 background fields. The latter have been linearly shifted to adjust for the differing Poisson noise level and the error bars indicate the formal 1$\sigma$ errors for 15 fields assuming that these are Gaussian (for the subsequent simulated backgrounds, this is not found to be the case).}\label{fig:src_bkgd_comp}
\end{figure}

\subsection{Constructing the Background Power Spectra}
We model the angular power due to these small-scale background fluctuations statistically as a multipole-dependent, log-normal additional component with means and standard deviations chosen to reproduce the difference in power and difference in the variances of the low- and high-resolution angular power spectra.  That is, we assume that the power in the high- and low-resolution angular power spectra, $\P_m^h$ and $P_m^l$, respectively, are related via
\begin{equation}
\P_m^h = \P_m^l + \P_m^{ss}
\end{equation}
where the small-scale structure contribution $\P_m^{ss}$ is a log-normal random variable, i.e., $\P_m^{ss}=\delta_m e^{\sigma_m X}$ for a zero-mean, unit-variance Gaussian random variable $X$.  The values of $\delta_m$ and $\sigma_m$ are fully defined by the means and variances of the $\P_m^h$ and $\P_m^l$; it is straightforward to show that choosing at each $m$
\begin{equation}
  \delta_m
  =
  \sqrt{ \frac{\Delta_m^4}{\Delta_m^2+\Sigma_m^2} },
  ~~
  \sigma_m^2
  =
  \ln\left(\frac{\Sigma_m^2+\Delta_m^2}{\Delta_m^2}\right),
\end{equation}
where $\Delta_m$ is the difference in the mean powers and $\Sigma_m^2$ is the difference in the variances at high and low resolution at a given $m$, uniquely reproduces the means and variance of $\P_m^h$ and $\P_m^l$.  This is done independently for angular power spectra constructed using Front-converted events, using Back-converted events, and using both Front- and Back-converted events, resulting in slightly different choices for $\delta_m$ and $\sigma_m$.  For the last case, the resulting $\delta_m$ are shown in Figure \ref{fig:ures} for the red and blue points. This is compared to the power spectra for the Galactic background of the 3FGL source subtracted \Fermi sky (black points).

In summary, we have modeled the background power with a combination of the contaminating 3FGL sources, a random realization of the unresolved completion of the 3FGL source population, the large scale ($>9^\circ$) gradients, and random realizations of the small-scale gradients.  At very high Galactic latitudes ($|b|>50^\circ$) this last component becomes insignificant, strongly implicating the Galactic contribution as its source.  The resulting background power spectra is shown in Fig.~\ref{fig:bkgd_sum}, subdivided into their origins.

\section{Results}\label{sec:4}

\begin{figure}
  \begin{center}
    \includegraphics[width=\columnwidth]{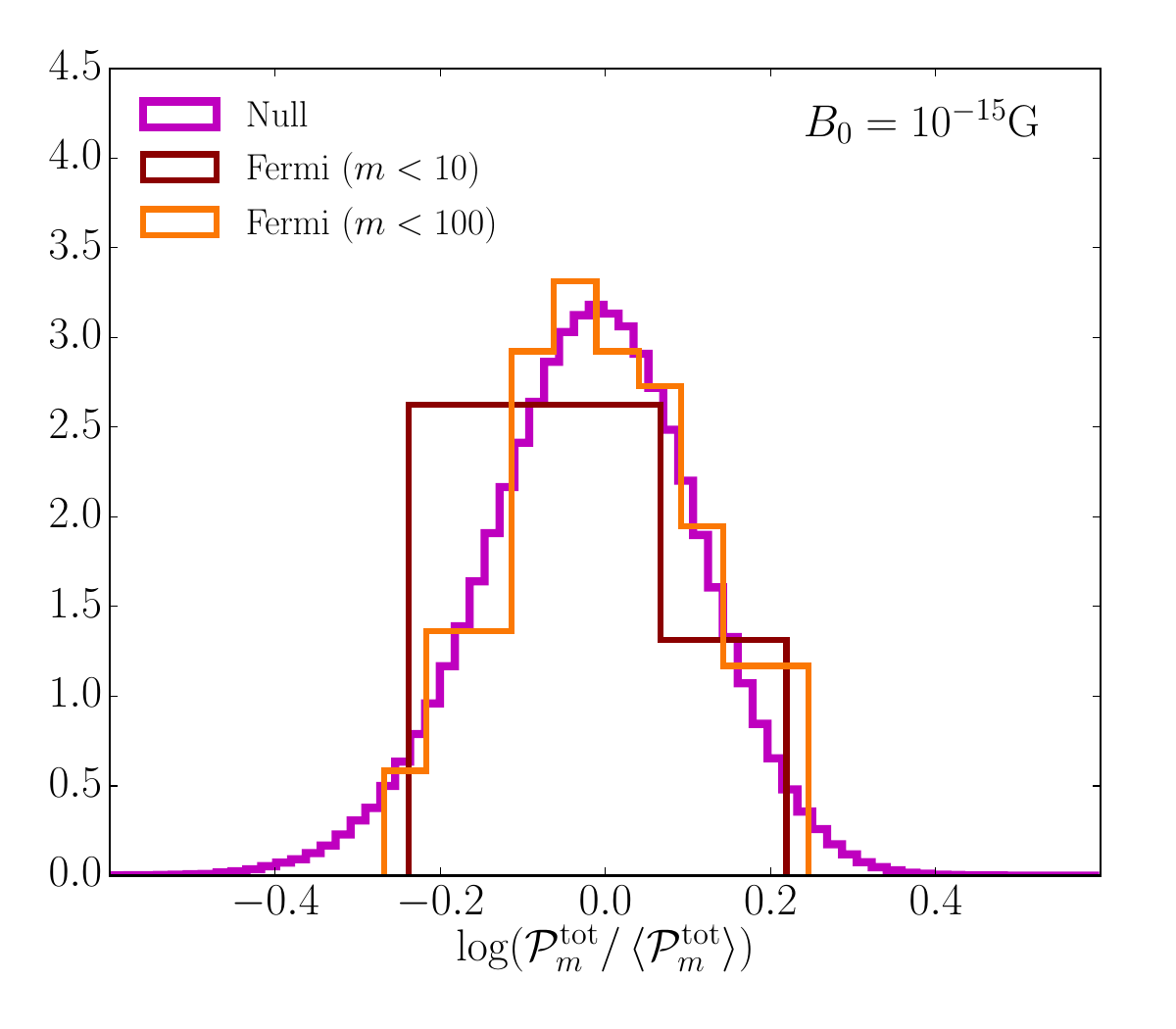}
  \end{center}
  \caption{Probability distribution of the observed power about the mean of the expected power for the $B_0=10^{-15}$~G optimized source sample for $m<10$ (brown) and $m<100$ (orange).  For reference, the distribution observed in the simulated realizations is also shown by the solid purple line.  The three are statistically indistinguishable.}\label{fig:dist}
\end{figure}

\begin{figure}
  \begin{center}
    \includegraphics[width=\columnwidth]{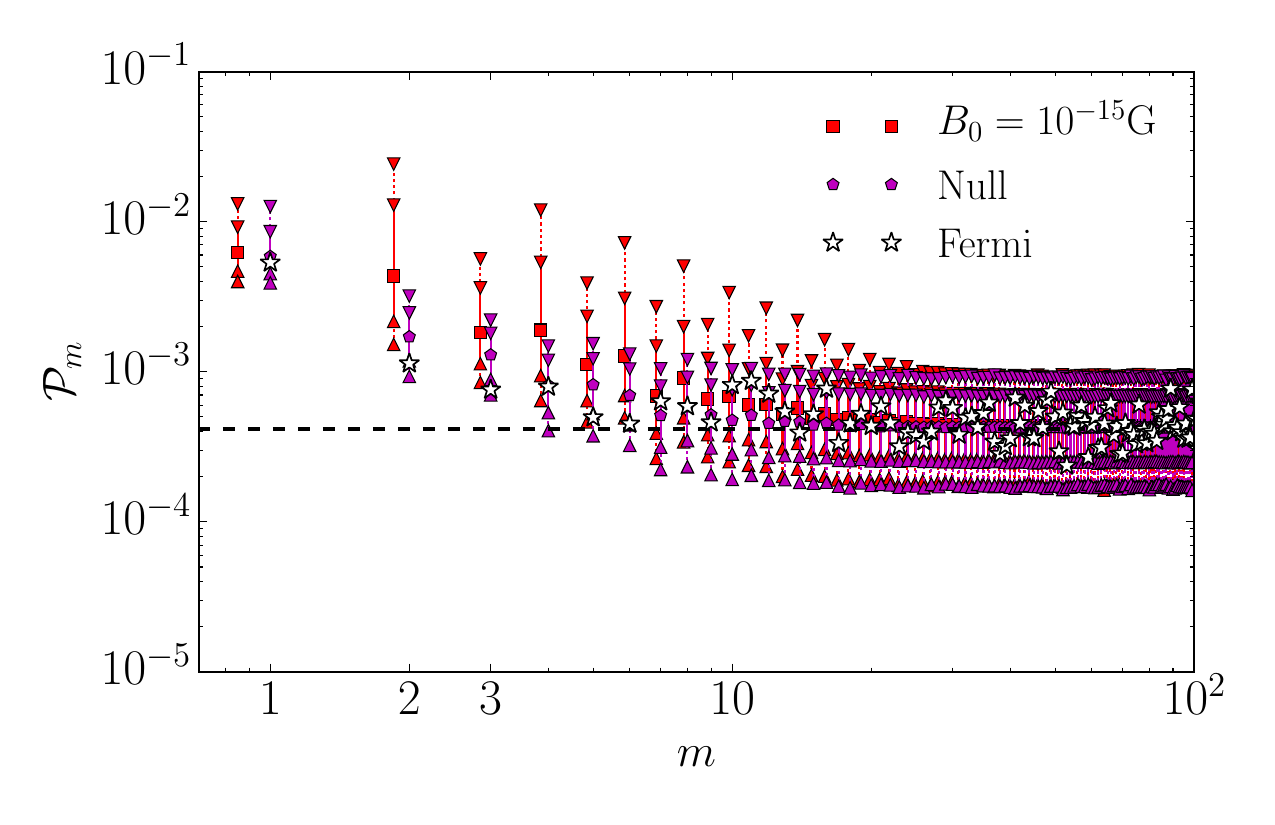}
    \includegraphics[width=\columnwidth]{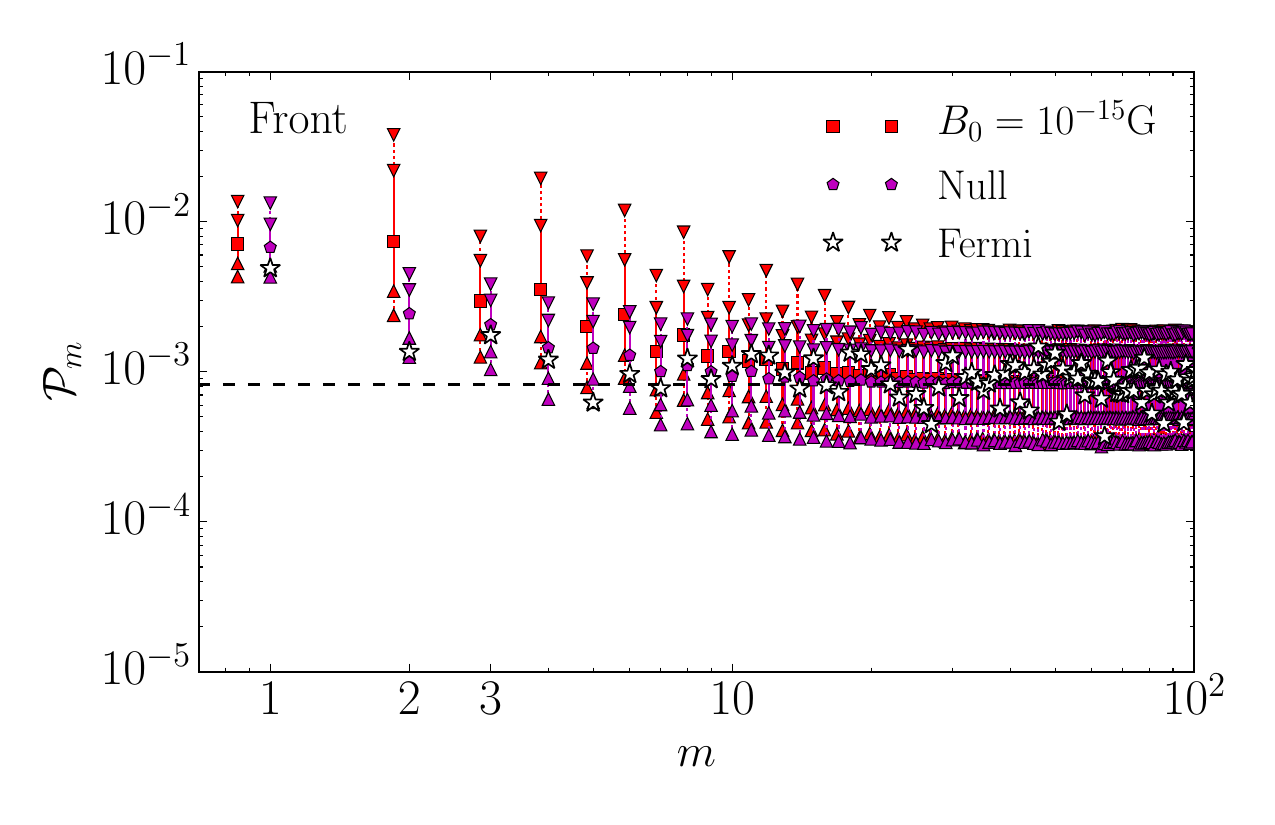}
    \includegraphics[width=\columnwidth]{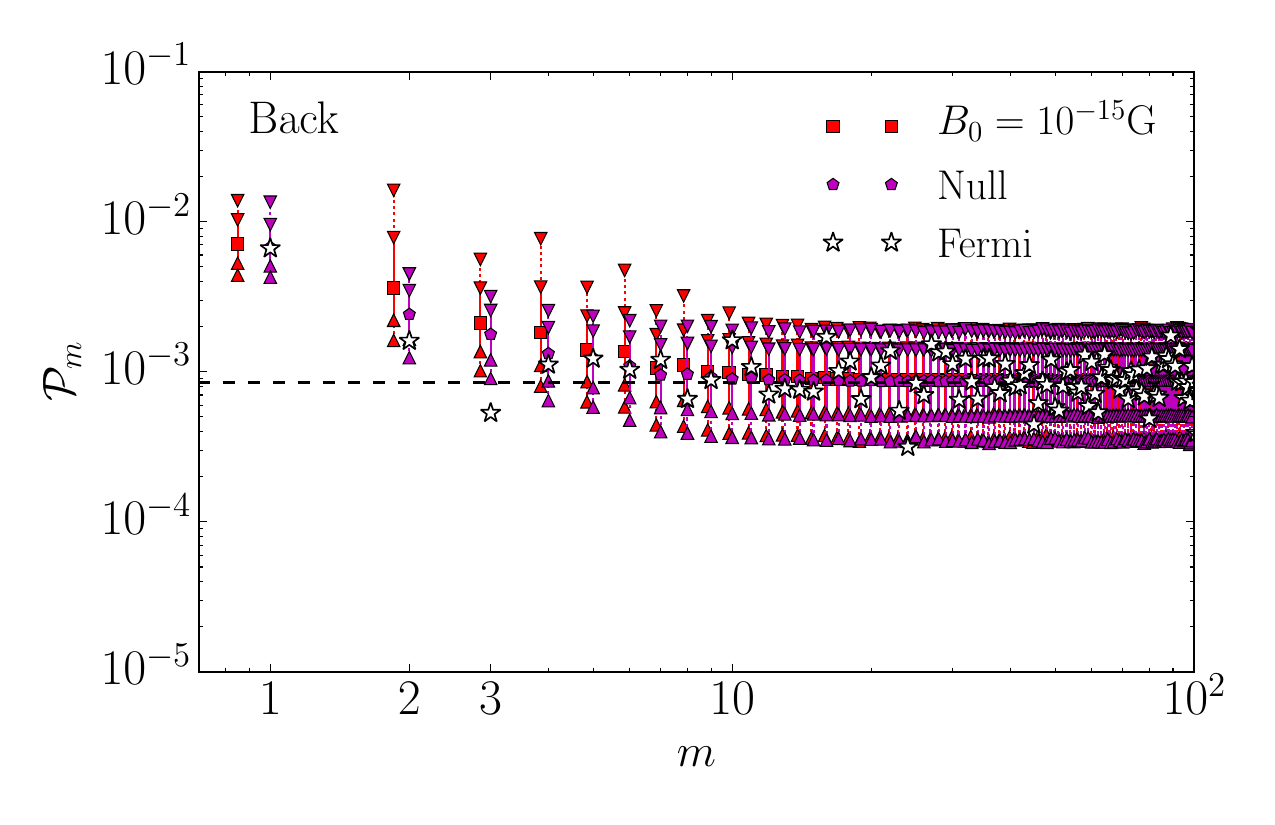}
  \end{center}
  \caption{Power spectra for gamma-ray halos with $B_0=10^{-15}$~G and the null case split into front (top) and back (bottom) detector contributions. The solid lines show the one-sided $95\%$ confidence intervals, while the dotted the $99\%$. No significant difference is observed between front and back detectors.}\label{fig:fbpscomp}
\end{figure}

\subsection{Comparing the power spectra to the background}
The stacked angular power spectrum for the optimized source set for $B_0=10^{-15}$~G is shown in Fig.~\ref{fig:src_bkgd_comp}.  As anticipated by the power spectrum from the background fields, there is excess power at low $m$.  However, this is entirely consistent with that expected for 15 background fields, adjusted via a constant shift for the different Poisson noise level (red points in Fig.~\ref{fig:src_bkgd_comp}).  Thus, there is no evidence for any excess power at low multipoles.

This is further evident in Fig.~\ref{fig:dist}, which shows the probability distribution of the simulated and observed power about the simulated median.  Not only is each value consistent with being drawn from the background, the distributions about the mean is itself similar.  Conversely, the ability to reproduce the statistics of the observed stacked angular power provides additional confidence in our ability to adequately model the background model.  This remains the case regardless of which optimized source sample is used.

Additionally, we expect different low $m$ to be correlated when pair halos exist \citep{BowTiesII}. While we have made no attempt to quantify the discrepancies, we found that the cross-correlations are consistent with the expected null distribution, and qualitatively look discrepant in the model that accounts for inverse-Compton halos (see \autoref{app:powcorr}).

\begin{figure*}[!t]
  \begin{center}
    \includegraphics[width=0.48\linewidth]{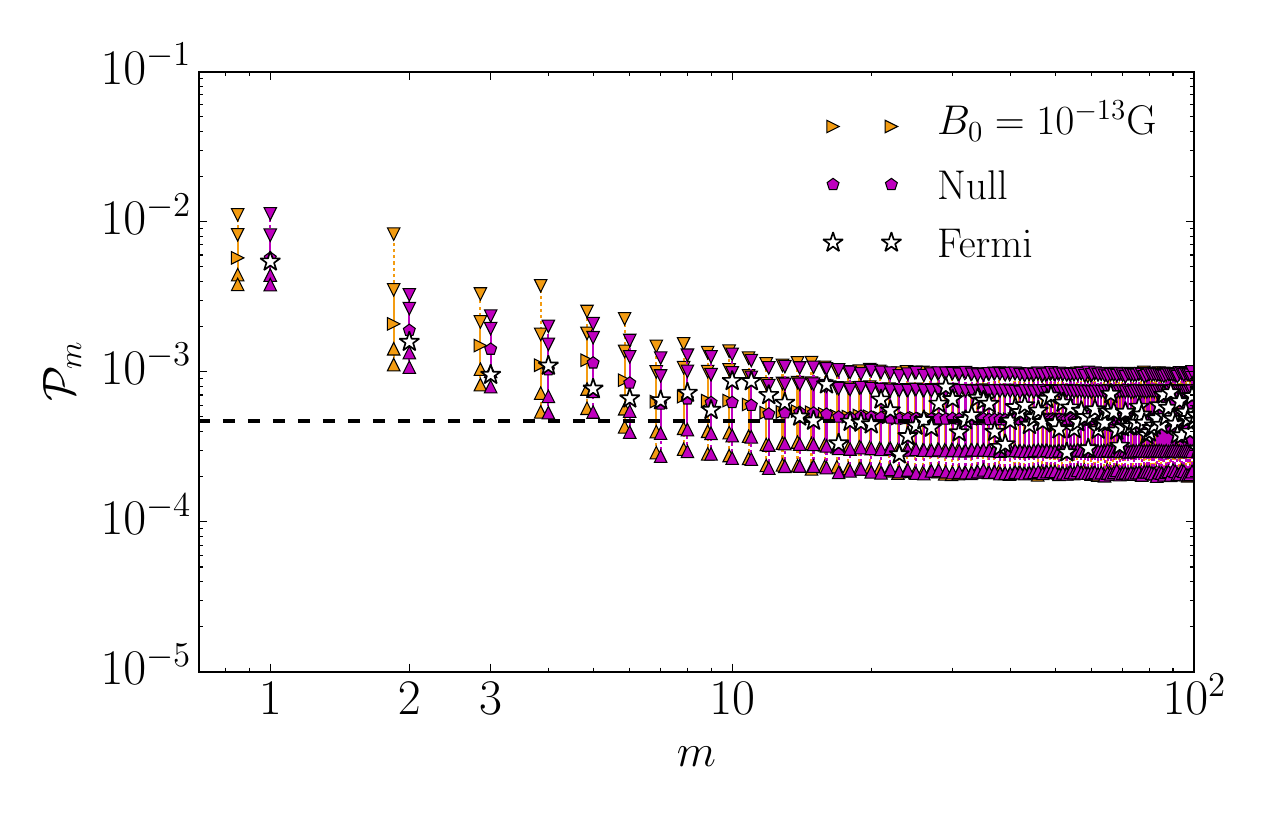}
    \includegraphics[width=0.48\linewidth]{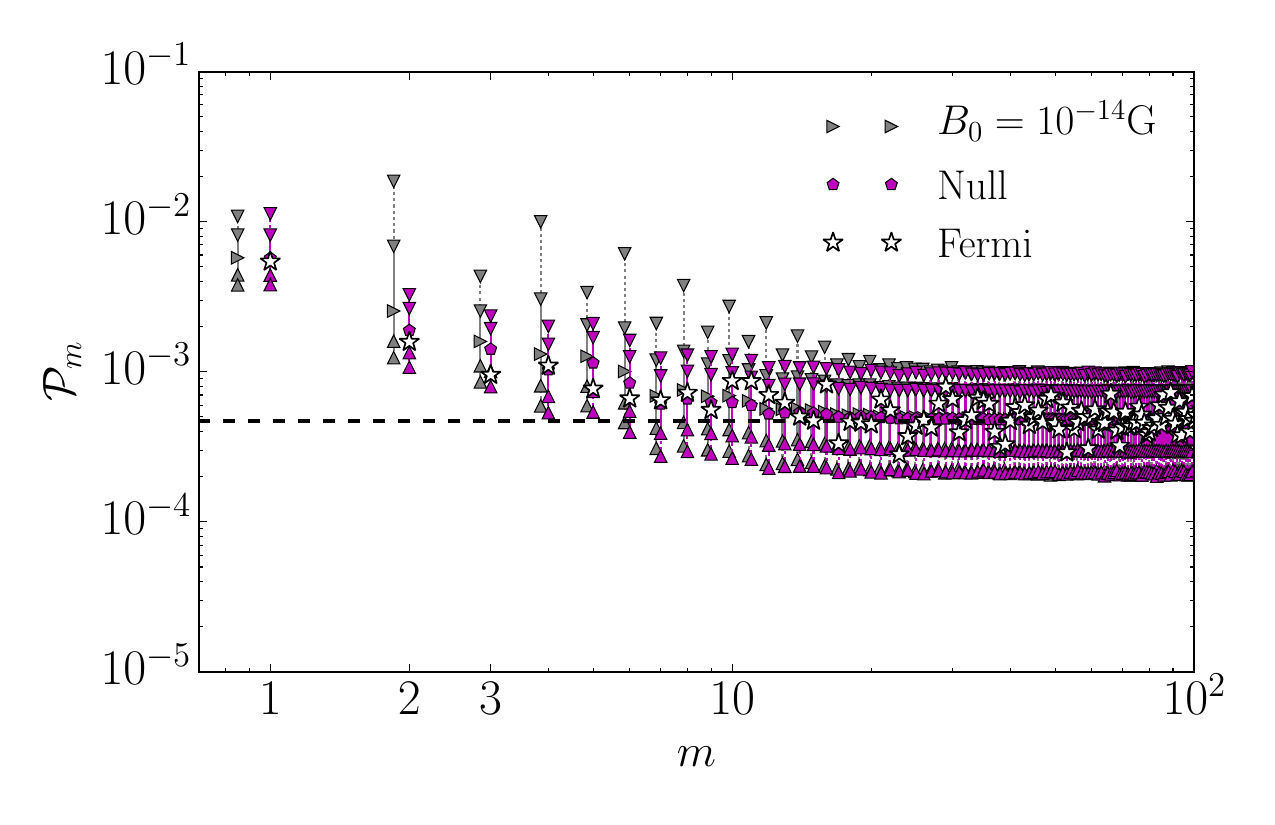}\\
    \includegraphics[width=0.48\linewidth]{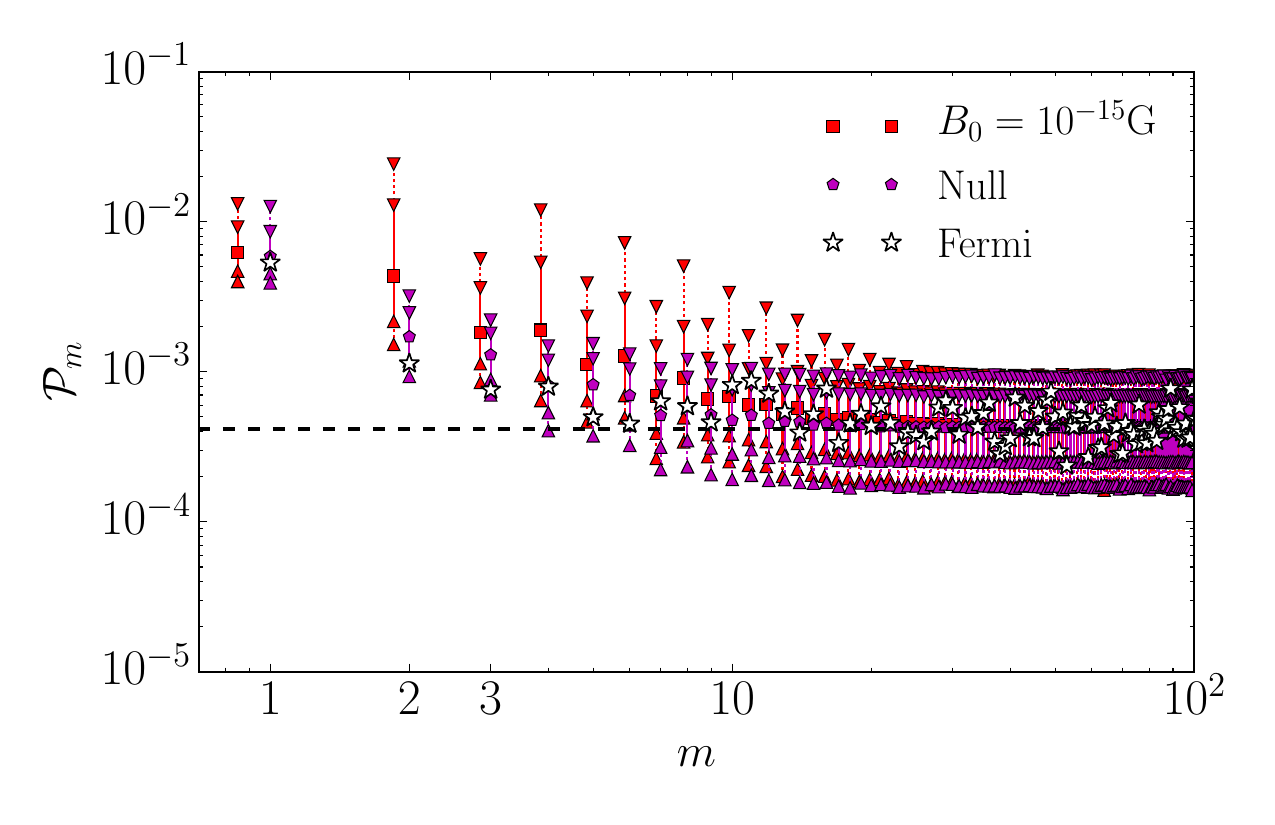}
    \includegraphics[width=0.48\linewidth]{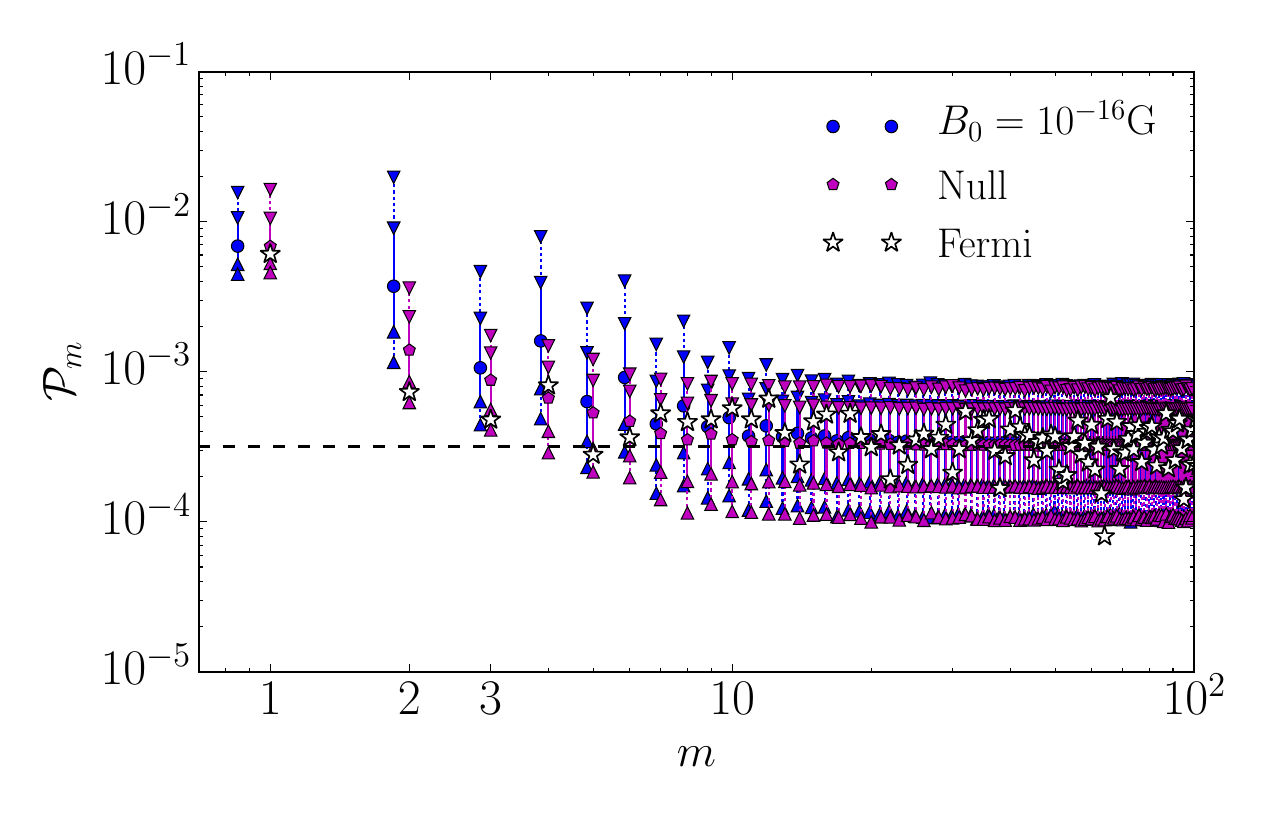}\\
    \includegraphics[width=0.48\linewidth]{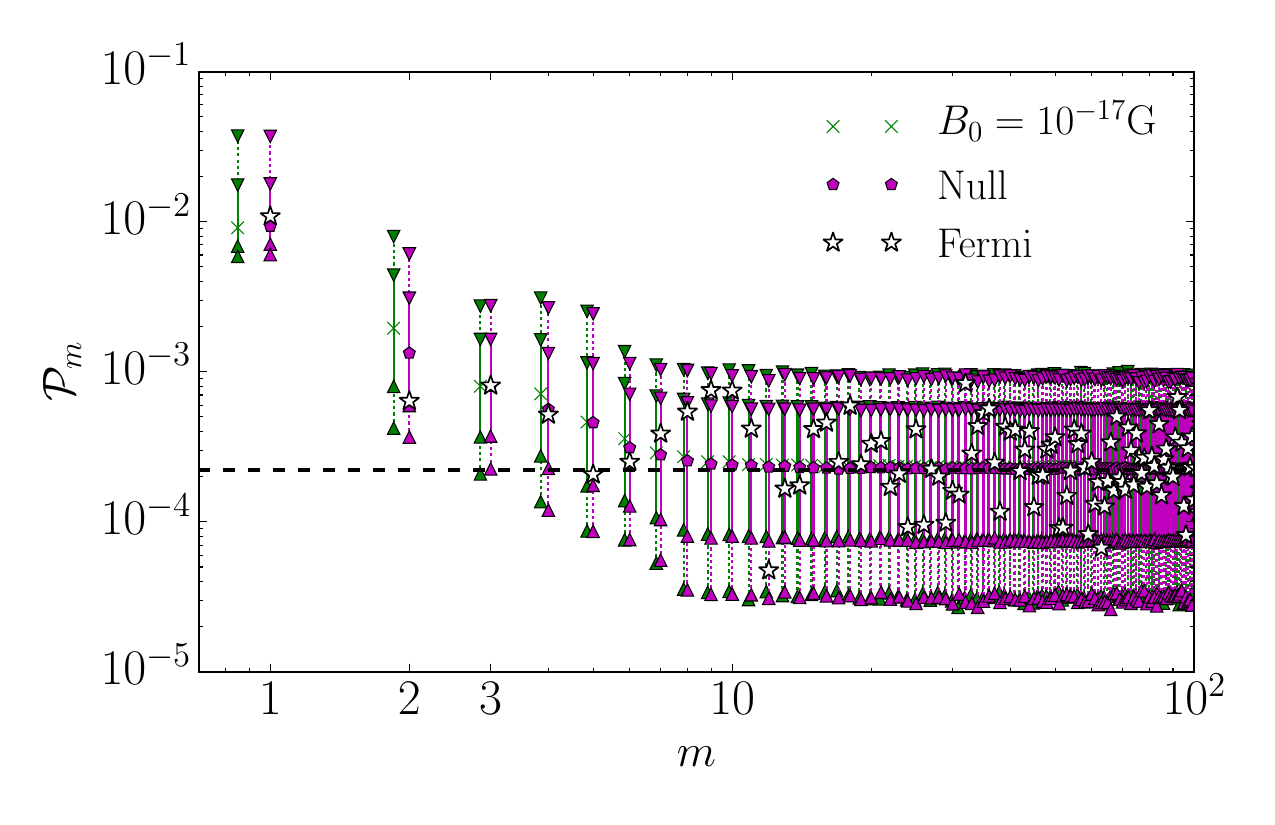}
    \includegraphics[width=0.48\linewidth]{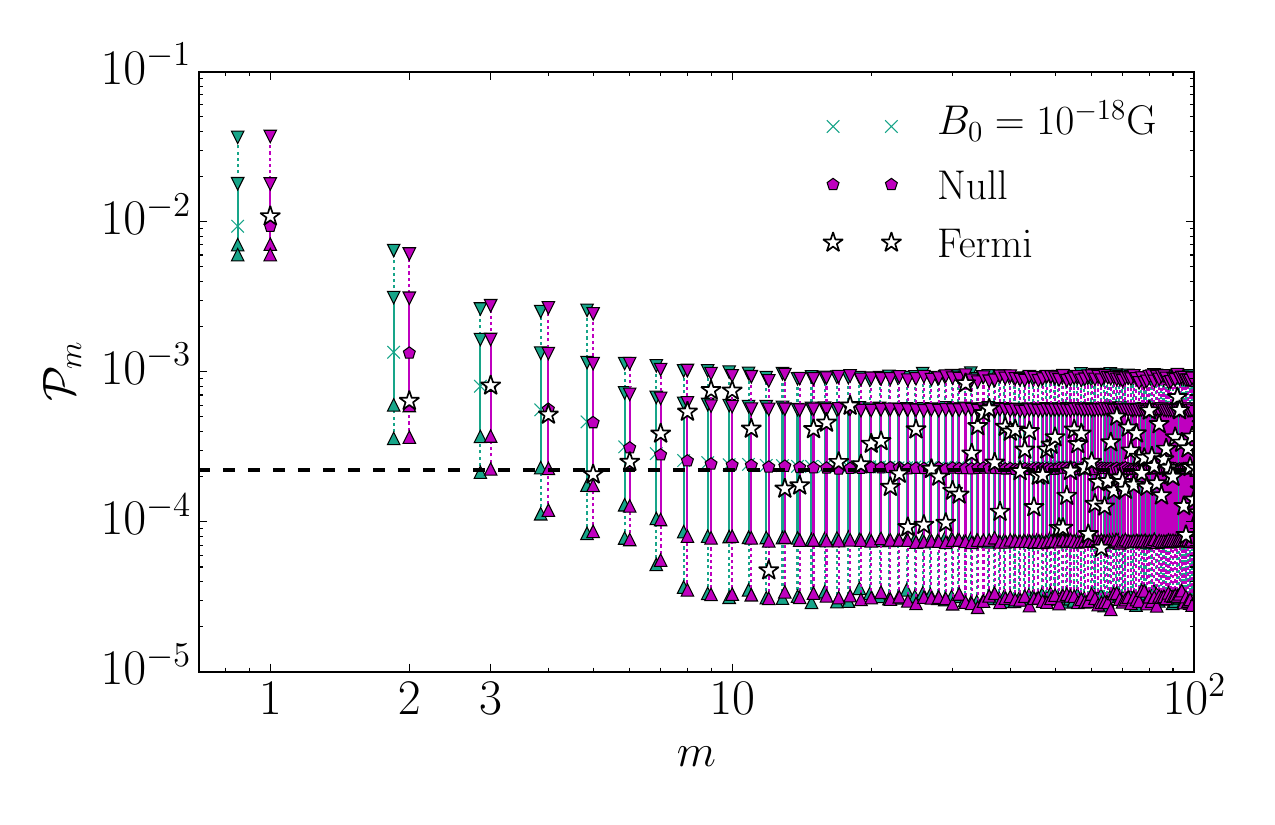}
  \end{center}
  \caption{Power spectra for halo models with the magnetic field strengths considered, i.e. $B_0=10^{-13}\G,~10^{-14}\G~,~10^{-15}\G10^{-16}\G,~10^{-17}\G,~10^{-18}\G$. In all instances the power spectra are consistent with the null hypothesis that no gamma-ray halos exist. The null power spectra are different in each panel since they use different source samples. For the source list see \citet{BowTiesII}.}\label{fig:otherBs}
\end{figure*}




\subsection{Quantifying the lack of gamma-ray halos}
To quantitatively assess the limits upon the IGMF strength that may be placed by the non-detection of excess structure, we employ our background model to generate multiple realizations of optimized \Fermi source lists, with and without inverse-Compton halos present.  From these we identify the one-sided confidence level at which a given value of the power may be excluded, i.e., the confidence level (CL), in each direction.  Because the stacked powers at different multipoles are strongly correlated in presence of inverse-Compton halos (see \autoref{app:powcorr}), we focus on the quadrupole exclusively when reporting constraints.  In principle, we can improve upon these by including information at multiple $m$; in practice, this serves to obscure the nature of the test for marginal benefit.  Note that because the optimized source lists vary with $B_0$, this procedure must be repeated for each IGMF strength.

In the top panel of Fig.~\ref{fig:otherBs} we show the 2$\sigma$ (95.5\%) and 3$\sigma$ (99.8\%) one-sided confidence levels for each multipole with (red) and without (purple) an inverse-Compton halo with $B_0=10^{-15}$~G. Even without modeling and including the background in the power spectra, we can exclude a large-scale IGMF with strength $B_0=10^{-15}$~G by 2$\sigma$.  When the background is included we can exclude our IGMF model at 3.9$\sigma$. Fig.~\ref{fig:fbpscomp} shows this result does not depend on which detector, front or back, is used, suggesting this is not from telescope systematics.

\begin{figure}
  \begin{center}
    \includegraphics[width=\columnwidth]{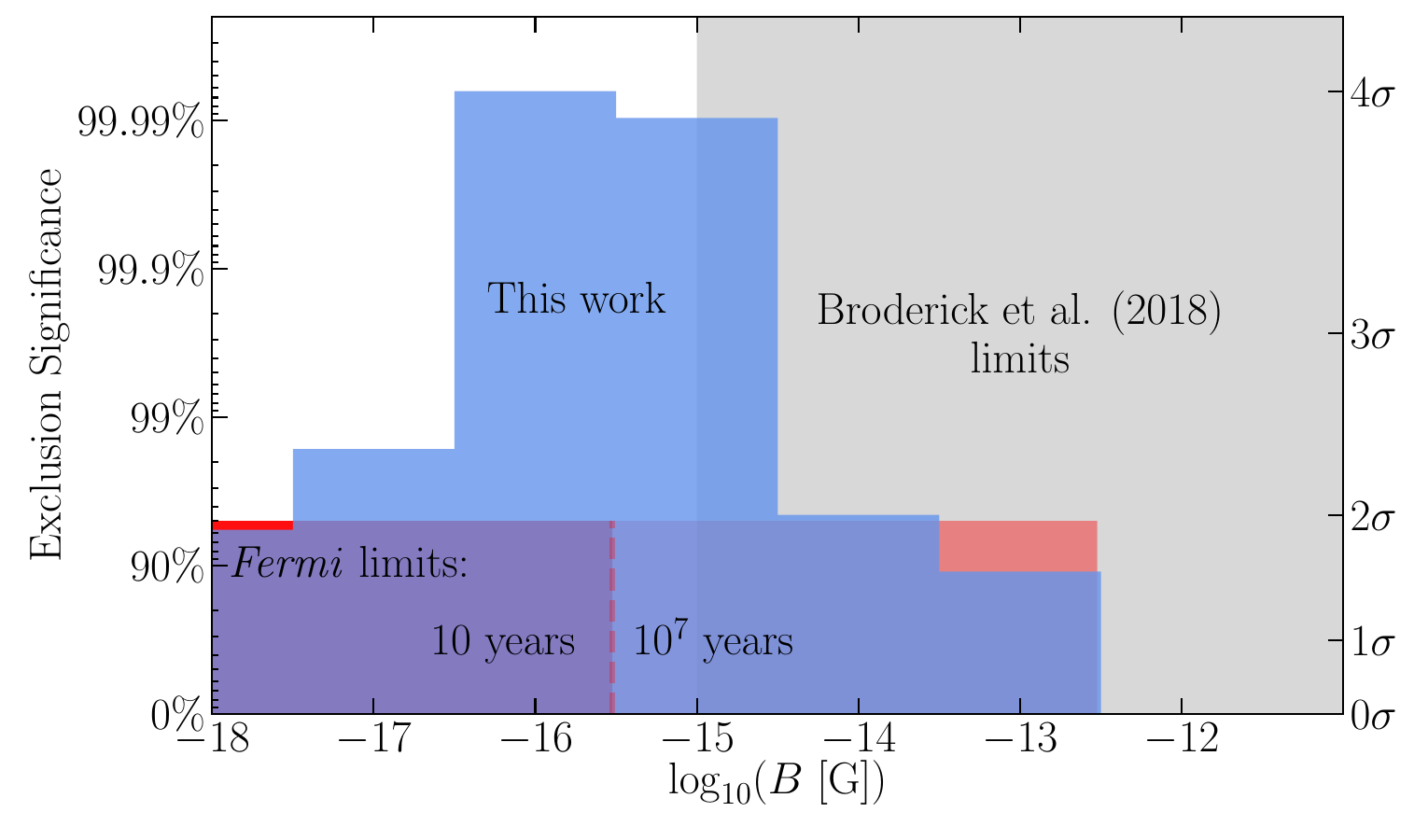}
  \end{center}
  \caption{Significance with which various large-scale ($\lambda_B>100$~Mpc) IGMF strengths may now be excluded.  The blue bars show the limits presented here.  The two red regions show the limits from \cite{Ackermann_2018}. The dark red assumes a duty cycle of 10 years while, the lighter red is for $10^7$ years. Above $10^{-15}~\G$ \cite{Broderick+2018} was able to exclude the existence of large-scale magnetic fields.}\label{fig:sigs}
\end{figure}

Moving to $B_0=10^{-18}$~G, $10^{-17}$~G, $10^{-16}$~G, $10^{-14}$~G, and $10^{-13}$~G, we generally find similar results. For $B_0=10^{-18}$~G and $10^{-13}$~G we used the optimized source list described in \cite{BowTiesI} for $B_0=10^{-17}$~G and $10^{-14}$~G, respectively (for their power spectra see Fig.~\ref{fig:otherBs}).  These may be excluded at 1.9$\sigma$, 2.4$\sigma$, 4.0$\sigma$, 2.0$\sigma$, and 1.6$\sigma$, respectively.  That is, for present-day field strengths between  $10^{-16}$~G and $10^{-15}$~G a large-scale IGMF can be excluded at more than 3.9$\sigma$.  At greater than 2$\sigma$ present-day field strengths slightly more than $10^{-18}$~G to $10^{-14}$~G a large-scale IGMF can be excluded.  These regions are shown in Fig.~\ref{fig:sigs} in comparison to the limits in \cite{Ackermann_2018} and \cite{Broderick+2018}.  Beyond those limits the inverse-Compton halos either become sufficiently small that they are confused with the source (small $B_0$) or extended that they lie beyond the regions about the sources considered (large $B_0$).

\section{Conclusions}\label{sec:5}
Combined with the SED lower limits, and making modest assumptions regarding the source duty cycles, we can exclude at more than 3.9$\sigma$ the existence of an IGMF of $10^{-16}<B_0<10^{-15}$~G, and at 2$\sigma$ field strengths of $10^{-17}<B_0<10^{-14}$, with $\lambda_B>100$~Mpc. The constraints presented in this work complement the results of \cite{Broderick+2018} and improve the limit of large-scale magnetic fields to be $<10^{-17}\G$ at 2$\sigma$. However, in this paper we used known gamma-ray bright sources, i.e. FSRQ and BL-Lac objects, from the \Fermi 3FGL source list.  As a result, while the constraints on the IGMF in this paper are statistically weaker than the \citet{Broderick+2018}, they require less assumptions about the nature of AGN and their high-energy emission.  Namely, the high-energy spectrum for each source is measured and we don't require AGN unification.  Therefore, these results form an independent test of the IGMF when compared to \citet{Broderick+2018} and is more direct.  Furthermore, we have demonstrated that even if the orientation of the gamma-ray halo is unknown i.e. the magnetic field orientation for each source is unknown, we can still constructively stack sources.  This work strongly constrains models of magnetogenesis that invoke inflationary processes in the early universe, which generate fields on large scales.

It is nevertheless worth pointing out that all of the above limits can be avoided if inverse-Compton cascades are preempted by other physical processes. Chief among these are large-scale beam plasma instabilities in the intergalactic medium.  The linear growth rates for these far exceed the inverse-Compton cooling times, a conclusion that appears to be robust even when higher order perturbative effects are included \citep{PaperI,Schl_etal:12,Chang:2014,Chang_etal:2016}. This may have important implications for the gamma-ray sky \citep{PaperV,PaperVa}, the thermal history of the intergalactic medium, and cosmic structure formation in low-density environments \citep{PaperII,PaperIII,PaperIV,Lamberts:2015}.  However, it does remain debated whether these instabilities saturate at sufficiently large amplitudes to efficiently extract the energy of the pairs in the nonlinear regime \citep{Mini-Elyv:12,sironi+giannios+2014,Chang:2014,Kempf+2016,Chang_etal:2016,resolution-paper,Vafin+2018,Vafin+2019}.

\acknowledgments

A.E.B., P.T. and M.S. receive financial support from the Perimeter Institute for Theoretical Physics and the Natural Sciences and Engineering Research Council of Canada through a Discovery Grant and Alexander Graham Bell scholarship (PT). Research at Perimeter Institute is supported by the Government of Canada through Industry Canada and by the Province of Ontario through the Ministry of Research and Innovation.  C.P., M.S., and E.P. acknowledge support by the European Research Council under ERC-CoG grant CRAGSMAN-646955. PC is supported by the NASA ATP program through NASA grant NNH17ZDA001N-ATP. AL acknowledges support by the Programme National des Hautes Energies (France). E.P. acknowledges support by the Kavli Foundation. 
\software{\texttt{astropy} \citep{astropy:2018}, \texttt{matplotlib} \citep{matplotlib}, \texttt{scipy} \citep{scipy}}

\bibliographystyle{aasjournal_aeb}
\bibliography{bigmh}

\appendix
\section{Power correlations}\label{app:powcorr}
\begin{figure}[!t]
  \begin{center}
    \includegraphics[width=0.48\columnwidth]{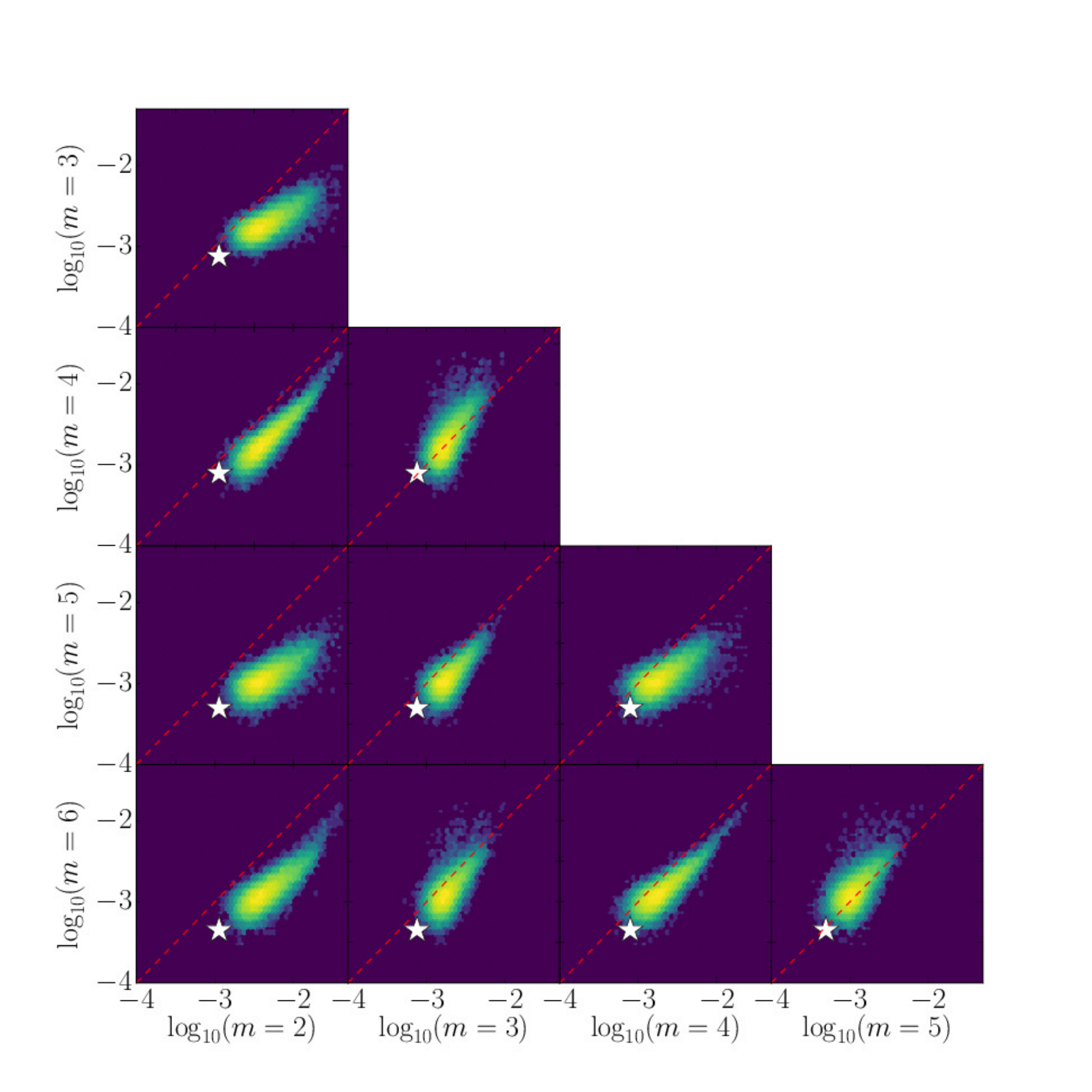}
    \includegraphics[width=0.48\columnwidth]{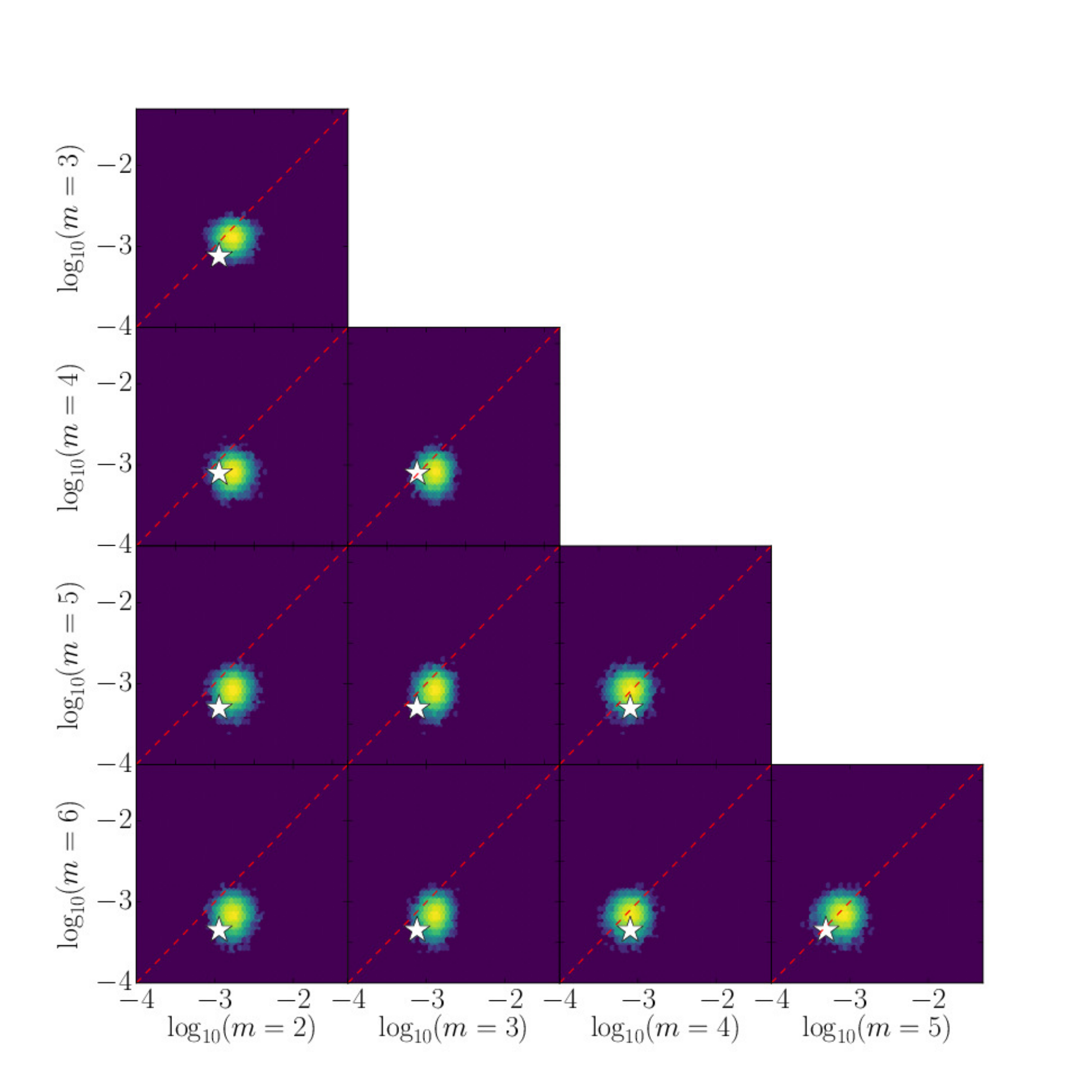}
  \end{center}
  \caption{Correlations in stacked angular power spectrum of the inverse-Compton halo (left) and Null (right) models (color map), in comparison to the observed values for a large-scale $10^{-15}$~G IGMF (white star).  In all panels the red dashed line shows the one-to-one relationship.} \label{fig:corr}
\end{figure}

Generally the powers at different multipoles are highly correlated when inverse-Compton halos are present \citep{BowTiesII}.  As a result, we have not made an attempt to leverage the observed broad inconsistency with the halo model expectation at low $m$.  Nevertheless, in Fig.~\ref{fig:corr} we show the joint probability distributions of the angular power in various low $m$ multipoles in comparison to the observed values.  From the comparison when inverse-Compton halos are present (Fig.~\ref{fig:corr}, left) it is clear that the observed powers lie well outside the joint distribution for all instances shown.  That is, the observations are even more highly inconsistent than implied by the quadrupolar comparison.  In contrast, when inverse-Compton halos are absent, i.e., our null case (Fig.~\ref{fig:corr}, right), the discrepancy disappears.

\end{document}